\documentclass[journal]{IEEEtran}
\usepackage[compatibility=false]{caption} 
\DeclareUnicodeCharacter{2061}{}
\usepackage{circuitikz} 
\usepackage{adjustbox}
\usepackage{algorithm}
\usepackage{algorithmic}
\usepackage{rotating}
\usepackage{tikz}
\usepackage{amsmath}
\usepackage{xcolor}
\usepackage{tikz}
\usetikzlibrary{shapes.geometric, arrows.meta, positioning}
\usepackage{xcolor}
\usepackage{float}
\usepackage{placeins}

\usepackage{tikz}
\usetikzlibrary{arrows.meta, positioning, shapes}
\usepackage{amsmath}
\usepackage{booktabs}
\usepackage{multirow}

\usepackage{cite}

\usepackage{graphicx}
\usepackage{subcaption}
\ifCLASSINFOpdf
\else
\fi
\usepackage{amssymb}
\usepackage{amsmath}
\usepackage{tabularx,booktabs}
\begin{document}
%
\title{Neural Network-Based Intelligent Reflecting Surface Assisted Direction of Arrival Estimation}
%
%
%


\author{Yasin~Azhdari\thanks{Y. Azhdari is with the Faculty of Electrical and Computer Engineering, Shiraz University, Shiraz, Iran (e-mail: yasin.azhdari@hafez.shirazu.ac.ir).}
	and Mahmoud~Farhang\thanks{M. Farhang is with the Faculty of Electrical and Computer Engineering, Shiraz University, Shiraz, Iran (e-mail: mfarhang@shirazu.ac.ir).}}
\maketitle

\begin{abstract}
Direction-of-Arrival (DoA) estimation assisted with an Intelligent Reflecting Surface (IRS) is crucial for various wireless applications, especially in challenging Non-Line-of-Sight (NLoS) environments. This paper presents a novel neural network-based architecture to address this challenge.

The key innovation is the introduction of a dedicated, learnable IRS layer integrated within a carefully designed end-to-end system established upon the physical and geometrical basis of the problem. Unlike conventional neural network layers, this specific one incorporates block diagonal sinusoidal weight constraints, where the phase arguments of these sinusoids are learned during training to directly emulate the phase shifts of the IRS elements. This allows the end-to-end system to optimize the IRS configuration for enhanced DoA estimation, eliminating the need for separate IRS optimization algorithms.  Moreover, different DoA regression networks, including a proposed structure, are presented and examined.

Numerical simulations, conducted under various conditions and noise levels, where controlled coherent multi-path components are introduced due to the presence of the IRS, demonstrate the superior performance of the novel end-to-end system compared to others and highlight its potential to significantly improve the accuracy of DoA estimation in complex IRS-assisted wireless systems. Besides, corresponding computational complexities of different approaches are also compared.  
\end{abstract}

\begin{IEEEkeywords}
Direction of arrival (DoA) estimation, Intelligent Reflecting Surface (IRS), Neural Network (NN), end-to-end system.
\end{IEEEkeywords}

%
\IEEEpeerreviewmaketitle

\section{Introduction}
%
%
%
%
\IEEEPARstart{E}{stimation}  of parameters corresponding to exponential signals contaminated in noise is a crucial problem in various signal processing applications, including array signal processing\cite{ye1995maximum}. Array signal processing aims to estimate parameters by exploiting both temporal and spatial information. Estimating the angular position of some sources by a set of sensors forming an array, known as Direction of Arrival (DoA) Estimation, is among main problems in the field of array signal processing. DoA estimation concerns determination of the angle of arrival of signals, in electromagnetic or acoustic wave forms, impinging on an array of antennas. DoA estimation has a variety of applications in wireless communications, sonar, radar, navigation and so on \cite{krim1996two,VanTrees_2008,chung2014doa,nielsen1991sonar,chen2010introduction,tuncer2009classical}.

Intelligent Reflecting Surfaces (IRSs), also known as Reconfigurable Intelligent Surfaces (RISs) have been applied in numerous fields in recent years, including wireless communications and radars. The main characteristic of these surfaces is that they have adjustable phase, amplitude, frequency and polarization, i.e., tunable electromagnetic response \cite{huang2019reconfigurable}. Non-line of sight paths are established in dead zones by proper exploit of an IRS and adjustment of its reflection coefficients, where line of sight links do not exist \cite{dardari2021nlos}.

An IRS-assisted approach is proposed to improve the SNR of received signals in order to enhance the performance of a detection system \cite{buzzi2021radar} by adjusting phases of the IRS. Moreover, the phase of the IRS is adjusted to maximize the SNR in the direction corresponding to the user equipment for localization in wireless communication under near-field propagation regime \cite{elzanaty2021reconfigurable}. Additionally, the IRS-assisted scheme is employed for DoA estimation problem, too. For example, in \cite{lan2022doa}, a Coprime Linear Array (CLA) is implemented by controlling IRS units. Also a DoA estimation method is presented corresponding to the IRS-based CLA.  In \cite{chen2022efficient}, a cost-effective direction-finding system using an Unmanned Aerial Vehicle (UAV) swarm is presented. This system includes a central full-functional receiving unit that performs DoA estimation by solely receiving the signals reflected by the IRS. Furthermore, \cite{chen2023doa} explores a method where the array receiver collects both the reflected signal from the IRS and the direct-path signal. By designing the phase of the IRS, the reflected path can be leveraged to enhance the accuracy of DoA estimation. Additionally, \cite{chen2023ris} formulates and addresses the DoA estimation problem in the presence of wireless communication interference using an IRS. An atomic norm-based approach is then proposed for joint DoA estimation and interference removal, where the optimization problem is also solved to design the matrix containing IRS characteristics for interference mitigation.

An important subject under IRS-related topics is how to adjust and tune its response, specifically the phase response, in order to attain the desired performance. In related works phase design is implemented to maximize the SNR towards the desired user equipment, to maximize the coherence of the signals from direct and reflected paths, to minimize the Cramer-Rao Lower Bound (CRLB) of the problem, or to remove the interference \cite{elzanaty2021reconfigurable, chen2023doa, song2023intelligent, chen2023ris}. Moreover, a reasonable approach is to employ Artificial Neural Networks (ANNs) to solve the associated optimization problems \cite{tariq2024deep}. For example, \cite{fathy2023machine} proposed machine-learning-based algorithms for the online implementation of active and passive beamforming to enhance energy efficiency for a IRS assisted single cellular network. However, to the best of our knowledge, no prior work has integrated a dedicated Neural Network (NN) layer as the IRS for DoA estimation–oriented phase design. This paper introduces a novel framework that embeds the IRS functionality within a dedicated NN layer, thereby directly tailoring the phase design to the DoA estimation task. Additionally, the idea of using a dedicated NN layer as the IRS can also be employed for other similar tasks.

An Artificial Neural Network (ANN) has shown the capability to establish the mapping between some input features (usually the raw measurements or the correlation matrix) and directions of sources. Both fully-connected  and convolutional neural structures with different modifications are used to address the problem of DoA estimation under variety of scenarios and have showed superior performance compared to the classical methods under certain conditions \cite{agatonovic2012application, fuchs2019single, liu2021doa, feintuch2023neural, liu2023attention, zheng2024deep}.

The main contribution of this paper is introduction of a novel neural network layer, which acts as the IRS (NN-based IRS layer) in a well-established end-to-end structure based on physical principles of the problem and providing its corresponding mathematics in order to learn the optimal phase design with respect to the task of DoA estimation, which has not been done yet to our best knowledge. Moreover, different DoA regression networks, including a proposed architecture, are used in the overall system. Also we will compare the performance of the proposed end-to-end system with the non-learning Maximum Likelihood (ML) approach. In this research, Uniform Planar Arrays (UPAs) are considered. But the method can be extended to other types of arrays.

The remainder of this paper is structured as follows: The problem model and formulation is described in Section 2. Section 3 reviews well-known traditional IRS assisted DoA estimation techniques and their corresponding algorithms. Section 4 presents the novel NN-based IRS layer and its corresponding mathematics, employed in well-thought-out end-to-end system for the task of DoA estimation. Numerical simulations and results are provided in Section 5. Finally, conclusions are given in Section 6.


\section{Problem Model}

An IRS-assisted scenario as shown in Fig. 1 is considered. According to figure 1, it is assumed that there is no line of sight path between the source and the antenna array. Thus the only path available is through IRS reflection. The antenna array and the IRS are both considered to be uniform planar arrays (UPAs).
The antenna array is placed on the yz plane of the coordinate system and the IRS is placed on the xy plane. The target is assumed to be in the far-field region with respect to the IRS and antenna, whereas the IRS is placed in the near-field region with respect to the antenna. Without loss of generality, one target at the non line of sight far-field region of the array is assumed, i.e., the most powerful one. Narrow-band waveform is assumed to prevent the effect of signal propagation on the IRS and the array. Also, unit gains and isotropic power receiving patterns are considered for both array antennas and IRS unit cells. As optimizing those factors are out of scope of this paper. 
\begin{figure}
	\centering
	\hspace*{5mm}
	\vspace*{-5mm}
	\includegraphics[origin=c,scale=0.50,trim={8cm 2cm 6cm 2cm},clip]{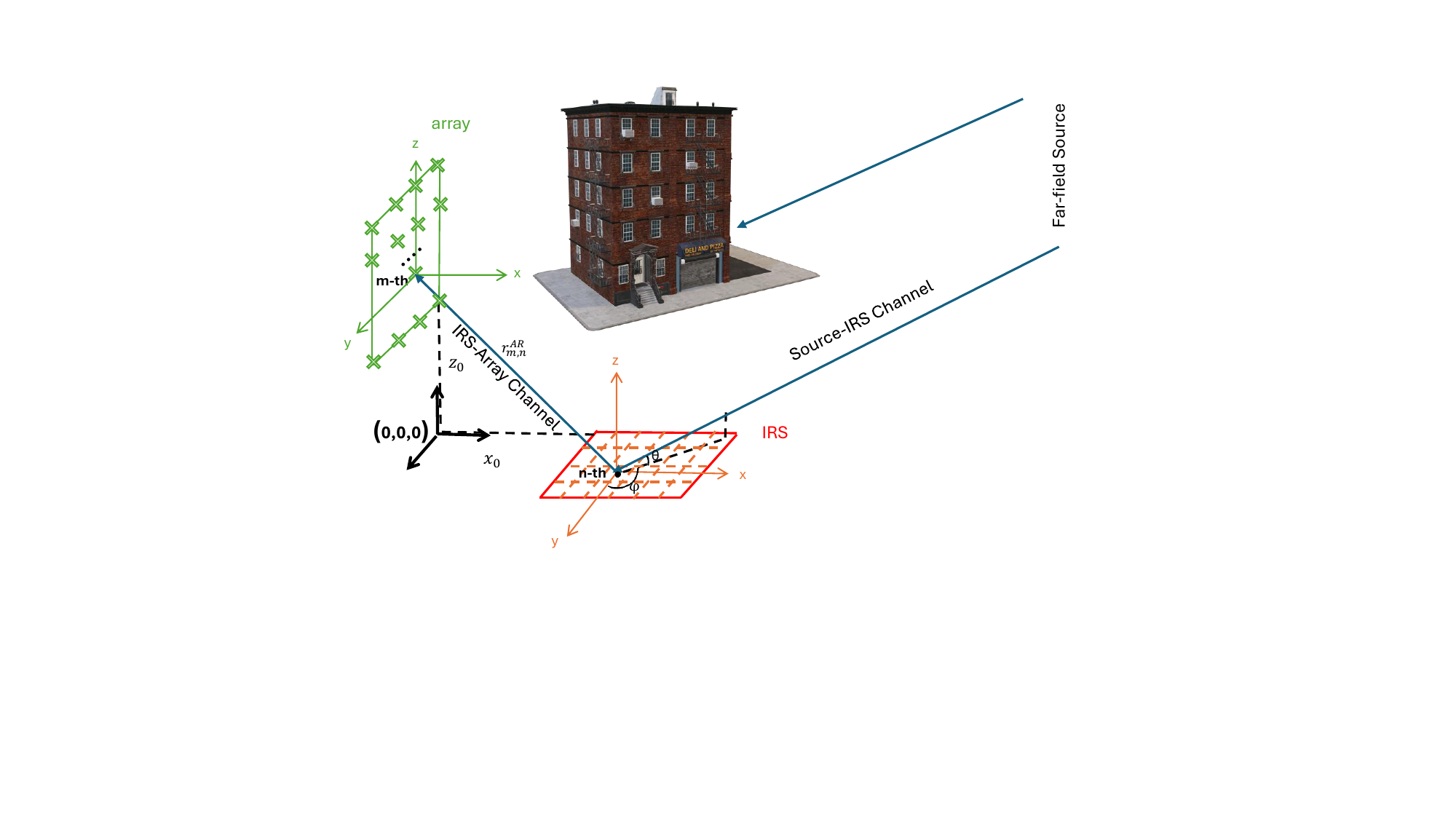}
	\caption{{Problem Model}}
	\label{fig1001}
\end{figure}

Let $M^A_y$ and $M^A_z$ denote the number of UPA elements along the $y$ and $z$ axes, with inter-element spacings of $d_y^A$ and $d_z^A$, respectively. Similarly, let $M^R_x$ and $M^R_y$ denote the number of IRS unit cells along the $x$ and $y$ axes, with inter-element spacings of $d_x^R$ and $d_y^R$. $M^A = M^A_y \times M^A_z$ and $M^R = M^R_x \times M^R_y$ denote the total number of ULA elements and IRS unit cells, respectively. The angle between the wavefront incident on the IRS and its projection on the xy plane is denoted as $\theta$. Moreover the angle between the projection on the xy plane and the y-axis is represented as $\phi$. {The position of the $m$-th UPA element with respect to the assumed coordinate center is given as:}
\begin{equation}
	\label{eqn_1000a}
	{p_{m}^A = (0, y_m^A, z_m^A)},
\end{equation}
{where $1 \leq m \leq M^A$.}
The position of the $n$-th IRS unit cell is:
\begin{equation}
	\label{eqn_1000b}
	{p_{n}^R = (x_n^R, y_n^R, 0)},
\end{equation}
{where $1 \leq n \leq M^R$.}

The path difference for the signal from the source to the IRS and from the RIS to the UPA is now computed considering the  spatial configuration. {The distance between the $m$-th UPA element and the $n$-th IRS unit cell is:}
\begin{equation}
	\label{eqn_1000c}
	{r_{m,n}^{AR} = \|p_{m}^A - p_{n}^R\|_2.}
\end{equation}

Moreover,
\begin{equation}
	\label{eqn_1000d}
	{r_{n}^{RT} = x_n^R \cos(\theta) \sin(\phi) + y_n^R \cos(\theta) \cos(\phi)}
\end{equation}
{denotes the path difference between the reference and the $n$-th IRS unit cell.}

Also $\gamma_n = |\gamma_n| e^{j\phi_n}$ denote the  adjustable reflection coefficient of the $n$-th unit cell. $|\gamma_n|$s are assumed to be 1 throughout this paper. Considering $\boldsymbol{\Phi}$ as the IRS phase shifts vector, $\boldsymbol{\omega} = e^{j\boldsymbol{\Phi}} = [e^{j\phi_1}, e^{j\phi_2}, . . . , e^{j\phi_{M^R}}]$ denotes the corresponding reflective steering vector. Then $\boldsymbol\Omega$ is defined as the $M^R \times M^R$ matrix containing the IRS phases as follows:

\begin{equation}
	\label{eqn_1001}
	\boldsymbol{\Omega} = \text{diag}(\boldsymbol{\omega}) = \text{diag}([e^{j\phi_1}, e^{j\phi_2}, \ldots , e^{j\phi_{M^R}}])
\end{equation}

Considering a far-field deterministic impinging source, the received signal from the reflection path can be modeled as  \cite{elzanaty2021reconfigurable, chen2023doa}:

\begin{equation}
	\label{eqn_1002}
	\mathbf{y_r}=(\mathbf{H}^{AR}\odot\mathbf{A}^{AR})\mathbf{\Omega}\mathbf{a}^{RT}s = \mathbf{a}_{r}s 
\end{equation}
where $\mathbf{A}^{AR}$ and $\mathbf{a}^{RT}$ are the array-IRS channel steering matrix and IRS-target channel steering vector, respectively.
\begin{equation}
	\label{eqn_1003}
	\mathbf{A}^{AR} = \left[\begin{array}{lr}
		e^{-j2\pi r_{1,1}^{AR}/\lambda} ... & e^{-j2\pi r_{1,M^R}^{AR}/\lambda} \\
		... & ...\\
		e^{-j2\pi r_{M^A,1}^{AR}/\lambda}... & e^{-j2\pi r_{M^A,M^R}^{AR}/\lambda} \end{array} \right] 
\end{equation}

\begin{equation}
	\label{eqn_1004}
	\mathbf{a}^{RT} = [e^{j2\pi r_{1}^{RT}/\lambda}, ... , e^{j2\pi r_{M^R}^{RT}/\lambda}]^T 
\end{equation}

Also, $\mathbf{H}^{AR}$ is the matrix representing the amplitude of the received signal from the near-field reflection paths \cite{tang2020wireless}
\begin{equation}
	\label{eqn_1002b}
	H_{m,n}^{AR} = \sqrt{\frac{P_r}{4\pi (r_{m,n}^{AR})^2}}
\end{equation}
where $P_r$ denotes the received power at the IRS which is assumed to be the same as the received power at the array due to the far-field assumption.

Thus the received signal at the array can be written as: 

\begin{equation}
	\label{eqn_1005}
	\mathbf{y}=\mathbf{y}_{r}+\mathbf{n}=(\mathbf{H}^{AR}\odot\mathbf{A}^{AR})\mathbf{\Omega}\mathbf{a}^{RT}s+\mathbf{n} = \mathbf{a}_{r}s+\mathbf{n} 
\end{equation}
where $\mathbf{n}$ is the additive noise distributed as $ CN(\mathbf{0} ,\sigma_n^2I)$.



The signal $s$ in the above formulation can also represent a vector containing multiple snapshots of the source signal. In this case, $\mathbf{s}$ has dimensions $L \times 1$, where $L$ is the number of snapshots. Consequently, the received signal $\mathbf{y}$ becomes:
\begin{equation}
	\label{eqn_1006}
	\mathbf{Y} = \mathbf{a}_r \mathbf{s} + \mathbf{N},
\end{equation}
where $\mathbf{a}_r$ denotes the steering vector corresponding to the NLoS path defined previously. $\mathbf{Y} \in \mathbb{C}^{M^A \times L}$ is the matrix of received signals, $\mathbf{s} \in \mathbb{C}^{1 \times L}$ is the vector of source signals, and $\mathbf{N} \in \mathbb{C}^{M^A \times L}$ represents the noise matrix.

\section{Classic IRS assisted DoA estimation }

\subsection{Phase adjustment method}
SNR Maximization \cite{elzanaty2021reconfigurable} and CRLB minimization \cite{song2023intelligent} are assumed for phase adjustment with the classic ML approach.

\subsubsection{SNR Maximization}
SNR  maximization criterion results in the following optimal designed phase for the $n$-th IRS unit cell \cite{elzanaty2021reconfigurable}:
\begin{equation}
	\label{eqn_1010}
	  \phi_{n}({\theta, \phi}) = \frac{2 \pi}{M^A \lambda}\sum_{m=1}^{M^A} (r_{m,n}^{AR} - r_n^{RT} + r_m^{AT})
\end{equation}
where,
\begin{equation}
	\label{eqn_1005d}
	{r_{m}^{AT} = y_m^A \cos(\theta) \cos(\phi) + z_m^A \sin(\theta)}
\end{equation}
{denotes the path difference between the reference and the $m$-th array element.} The coarse estimates of corresponding angles are leveraged to obtain the phase shifts.
\subsubsection{CRLB Minimization}
Minimizing the CRLB \cite{song2023intelligent} by selecting the proper set of phases $\mathbf{\Phi}$ in our specific problem model is considered as in \cite{chen2023doa}:
\begin{equation}
	\label{eqn_1023}
	\text{min}_\mathbf{\omega} \; \text{CRLB}_{\theta}(\mathbf{\omega}) + \text{CRLB}_{\phi}(\mathbf{\omega})\;\; \text{s.t.} \;\; |\omega_i|=1 \; \forall  i=1, ..., M^R
\end{equation}
where $\text{CRLB}_{\theta}(\omega)$ and  $\text{CRLB}_{\phi}(\omega)$ for our specific problem model and formulation, can be obtained in a similar fashion to \cite{stoica1989music} (signal and noise variance are assumed to be known) :
\begin{equation}
	\label{eqn_1008}
	\text{CRLB}_{x} = \frac{\sigma_n^2}{2\sigma_s^2} \text{Re}\{\frac{\partial\mathbf{a}}{\partial x}^H (I-\mathbf{a}(\mathbf{a}^H\mathbf{a})^{-1}\mathbf{a}^H)\frac{\partial\mathbf{a}}{\partial x}\}^{-1}
\end{equation}
where $x= \theta$, or $\phi$ and $\sigma_s^2 = \text{E}\{|s|^2\} = |s|^2$.
Since the CRLB function and the unit-modulus constraints are not convex, The optimization problem of the IRS phase design is not convex and thus is challenging and complicated. In order to transform this problem into a convex optimization problem, the semi definite relaxation (SDR) and successive convex approximation (SCA) techniques can be employed \cite{song2023intelligent}. Also the method of Riemannian manifold optimization can be applied for designing optimal phase shifts \cite{chen2023doa}. Here we deal with complex circle manifold, and to solve the corresponding Riemannian manifold-based optimization problem (\ref{eqn_1023}) the Riemannian Steepest Descent (SD), Conjugate Gradient (CG), and Trust-Region (TR) methods are proposed before \cite{boumal2023intromanifolds}. 
The TR method is used through this paper.

\subsection{Maximum Likelihood Doa Estimation}
The Maximum Likelihood (ML) DoA estimator for a deterministic source signal is implemented as follows and requires a grid search \cite{boman2001low}:

\begin{equation}
	\label{eqn_1019}
	\hat{\theta}, \hat{\phi} = \textit{argmax}_{\theta , \phi} \; \lambda_{max}[(\mathbf{a}^H\mathbf{a})^{-\frac{1}{2}}(\mathbf{a}^H \mathbf{yy}^H \mathbf{a})(\mathbf{a}^H\mathbf{a})^{-\frac{1}{2}}]
\end{equation}
where $\lambda_{max}[\mathbf{M}]$ represents the maximum eigenvalue of the matrix $\mathbf{M}$.

\section{Learning-based IRS assisted DoA estimation}
As previously discussed, various non-learning optimization methods have been proposed to adjust the phases of IRS units. In this paper, we propose a novel approach by implementing the IRS using a dedicated neural network layer. This layer is designed to learn phase adjustments specifically tailored for the task of Direction of Arrival (DoA) estimation. The corresponding mathematics, and the end-to-end system structure are presented.

During training, the end-to-end system takes model-based simulated measurements observed at the IRS as inputs in order to optimize corresponding IRS phases respecting the task of DoA estimation, performed by the DoA regression network. Since neural networks typically expect real-valued inputs, the complex-valued measurements from each IRS unit are preprocessed by separating their real and imaginary components and interleaving them. These processed measurements are then fed into the IRS layer.
During test, observations at the array, based on optimized IRS phase configurations and affected by AWGN, i.e., the received signal, is used for DoA estimation. 

\subsection{Proposed IRS Layer Mechanism}
Forward propagation (FP) and backward propagation (BP) form the backbone of training and optimization in neural networks. 



Unlike common neural network blocks with tunable constant weights, the IRS block requires weights that are sinusoidal in nature, where the arguments (or phases) of these weights must be tuned during the training process in order to optimize the IRS phases. Consequently, both the forward propagation (FP) and backward propagation (BP) procedures must be adapted to accommodate this unique weight structure. 

The following subsections detail the process.
\subsubsection{Mathematical Foundation}
Under NLoS scenario, a complex-valued element of the IRS observation vector can be represented as:  $x_i = a_i + jb_i$ for $i = 1, \dots, M^R$, where $a_i$ and $b_i$ denote the real and imaginary components, respectively. Multiplying this observation by a complex exponential $e^{j\phi_i}$ results in the following expression per each element:
\begin{equation}
	x_i e^{j\phi_i} = \big(a_i \cos(\phi_i) - b_i \sin(\phi_i)\big) + j\big(a_i \sin(\phi_i) + b_i \cos(\phi_i)\big).
\end{equation}

Separating the real and imaginary components of the result and interleaving them for each element as follows, this operation can be expressed in the matrix form.

\begin{equation}
	\mathbf{x} = [a_1, b_1, a_2, b_2, \dots, a_{M^R}, b_{M^R}]^T.
\end{equation}

The result is then achieved through applying a block diagonal matrix $\mathbf{W}$ to the $\mathbf{z}$, where each $2 \times 2$ block $\mathbf{W}_i$ is parameterized by the phase angle $\phi_i$:
\begin{equation}
	\mathbf{W}_i = \begin{bmatrix}
		\cos(\phi_i) & \sin(\phi_i) \\
		-\sin(\phi_i) & \cos(\phi_i)
	\end{bmatrix}.
\end{equation}

The complete block diagonal weight matrix $\mathbf{W}$ is then represented as:
\begin{equation}
	\mathbf{W} = \text{diag}(\mathbf{W}_1, \mathbf{W}_2, \dots, \mathbf{W}_{M^R}).
\end{equation}

The transformed vector $\mathbf{z}$, which includes the interleaved real and imaginary parts of the desired result, is then obtained as:
\begin{equation}
	\mathbf{z} = \mathbf{W} \mathbf{x}.
\end{equation}

This formulation reveals that the multiplication of each complex observation $x_i = a_i + jb_i$ by $e^{j\phi_i}$ for $i = 1, \dots, M^R$  is equivalent to applying a block diagonal transformation matrix $\mathbf{W}$ in the real domain.

\subsubsection{Definition of the Weight Matrix}
The proposed IRS layer employs the weight matrix $\mathbf{W}$ described in (20), consisting of ${M^R}$ trainable $2 \times 2$ blocks, described in the previous part. Each block is parameterized by a single variable $\phi_i$, corresponding to a phase adjustment.

The parameters $\phi_i$ for $i = 1, \dots, {M^R}$ are initialized uniformly within $[-\pi, \pi]$ and updated during training to optimize the phase adjustments for the observations.

\subsubsection{Forward Propagationand and Backward Propagation for the IRS Layer}
In the forward pass, the IRS layer applies the block diagonal transformation $\mathbf{W}$ to the input vector $\mathbf{x}$, which consists of the interleaved real and imaginary components of the IRS observations. The output of the layer, which has the same dimension as input, is formulated as follows (no biases are considered for this layer and it applies no activation function):
\begin{equation}
	\mathbf{z} = \mathbf{W} \mathbf{x}.
\end{equation}

The backward pass involves calculating the gradients of the loss function $E$ with respect to the trainable parameters $\phi_i$ and the inputs $\mathbf{x}$. 

Each block $\mathbf{W}_i$ is parameterized by $\phi_i$, and the derivative of $\mathbf{W}_i$ with respect to $\phi_i$ is:
\begin{equation}
	\frac{\partial \mathbf{W}_i}{\partial \phi_i} = \begin{bmatrix}
		-\sin(\phi_i) & \cos(\phi_i) \\
		-\cos(\phi_i) & -\sin(\phi_i)
	\end{bmatrix}.
\end{equation}

Since $\mathbf{z} = \mathbf{W}\mathbf{x}$, and $\mathbf{W}$ is block diagonal, only the elements $z_{2i-1}$ and $z_{2i}$ depend on $\phi_i$. Therefore:

\begin{align}
	\frac{\partial z_{2i-1}}{\partial \phi_i} &= \begin{bmatrix} 1 & 0 \end{bmatrix} \frac{\partial \mathbf{W}_i}{\partial \phi_i} \begin{bmatrix} x_{2i-1} \\ x_{2i} \end{bmatrix} \nonumber \\
	&= -x_{2i-1}\sin(\phi_i) + x_{2i}\cos(\phi_i)\\
	\frac{\partial z_{2i}}{\partial \phi_i} &= \begin{bmatrix} 0 & 1 \end{bmatrix} \frac{\partial \mathbf{W}_i}{\partial \phi_i} \begin{bmatrix} x_{2i-1} \\ x_{2i} \end{bmatrix} \nonumber \\
	&= -x_{2i-1}\cos(\phi_i) - x_{2i}\sin(\phi_i)
\end{align}

Using the chain rule, the gradient of the loss with respect to $\phi_i$ is:
\begin{equation}
	\frac{\partial E}{\partial \phi_i} = \frac{\partial E}{\partial z_{2i-1}} \frac{\partial z_{2i-1}}{\partial \phi_i} + \frac{\partial E}{\partial z_{2i}} \frac{\partial z_{2i}}{\partial \phi_i}.
\end{equation}

Moreover, the gradient of the loss with respect to the inputs $\mathbf{x}$ is computed as:
\begin{equation}
	\frac{\partial E}{\partial \mathbf{x}} = \mathbf{W}^T \frac{\partial E}{\partial \mathbf{z}}.
\end{equation}

The trainable parameters $\omega_i$ are then updated using gradient descent:
\begin{equation}
	\phi_i \gets \phi_i - \eta \cdot \frac{\partial E}{\partial \phi_i},
\end{equation}
where $\eta$ is the learning rate. This procedure ensures convergence to optimal phase adjustments during the training.

The Mean Squared Error (MSE) is employed as the loss function for the DoA regression task:
\begin{equation}
	E = \frac{1}{n} \sum_{i=1}^n \big(z_i - \hat{z}_i\big)^2,
\end{equation}
where $\hat{z}_i$ and $z_i$ are the ground truth and predicted outputs, respectively. Besides, the gradient of the loss with respect to $\mathbf{z}$ is:
\begin{equation}
	\frac{\partial E}{\partial \mathbf{z}} = \frac{2}{n} (\mathbf{z} - \hat{\mathbf{z}}).
\end{equation}

All in all, 1) Unlike conventional layers, the proposed IRS layer explicitly models the phase-shifting behavior of IRS-aided systems using a trainable block diagonal matrix. This is a unique contribution to the field.
2) By parameterizing each 2$\times$2 block with a single phase variable $\omega_i$, the IRS layer achieves a compact representation with minimal computational overhead.
3) The sinusoidal structure of the weight matrix ensures smooth gradients, facilitating effective training using backpropagation.
4) The layer is fully consistent with the physical principles of the IRS-aided wireless communication system.

These features underscore the innovation and applicability of the proposed IRS layer in end-to-end learning-based architectures for DoA estimation and also other parameter estimation tasks.

\subsection{Implementation of the end-to-end system}
As discussed before, observations at IRS unit cells are used as input for training. Thus in order to formulate the input of the system during training,  we consider the target-IRS link as a direct path and simulate IRS observations as following:

\begin{equation}
	\label{eqn_1005_b}
		\mathbf{r} = \mathbf{a}^{RT}\mathbf{s}
\end{equation}
where $\mathbf{r}$ denotes observations at IRS unit cells. Also, $\mathbf{a}^{RT}$ represents the IRS-target channel steering vector introduced before.

Next, the real and imaginary components of observation of each unit are extracted. A new observation vector ($\mathbf{x}$) is then constructed by interleaving the imaginary components with their respective real counterparts. 


The IRS layer is then applied. Following the IRS layer, a fixed weight layer is employed to implement the mapping of observations at the IRS (after applying the corresponding phases) to observations at the array using the fixed weight matrix ($\mathbf{H}^{AR}\odot\mathbf{A}^{AR}$) as defined in equation (\ref{eqn_1002}). This mapping is determined based on the geometric configuration of the problem, which is assumed to remain constant throughout. Then in the same trend as \cite{o2017introduction, 8792076}, the channel AWGN with corresponding SNR randomly selected from the set \{$-20$, $-10$, 0, 10,20\} dB, is applied during training. 
 
\begin{figure}[h]
	\centering
	\begin{subfigure}[b]{0.2\textwidth}
	\centering
	
	\hspace*{-25mm}
	\vspace*{0mm}
	\includegraphics[scale=0.3,trim={0cm 0cm 0cm 4cm},clip]{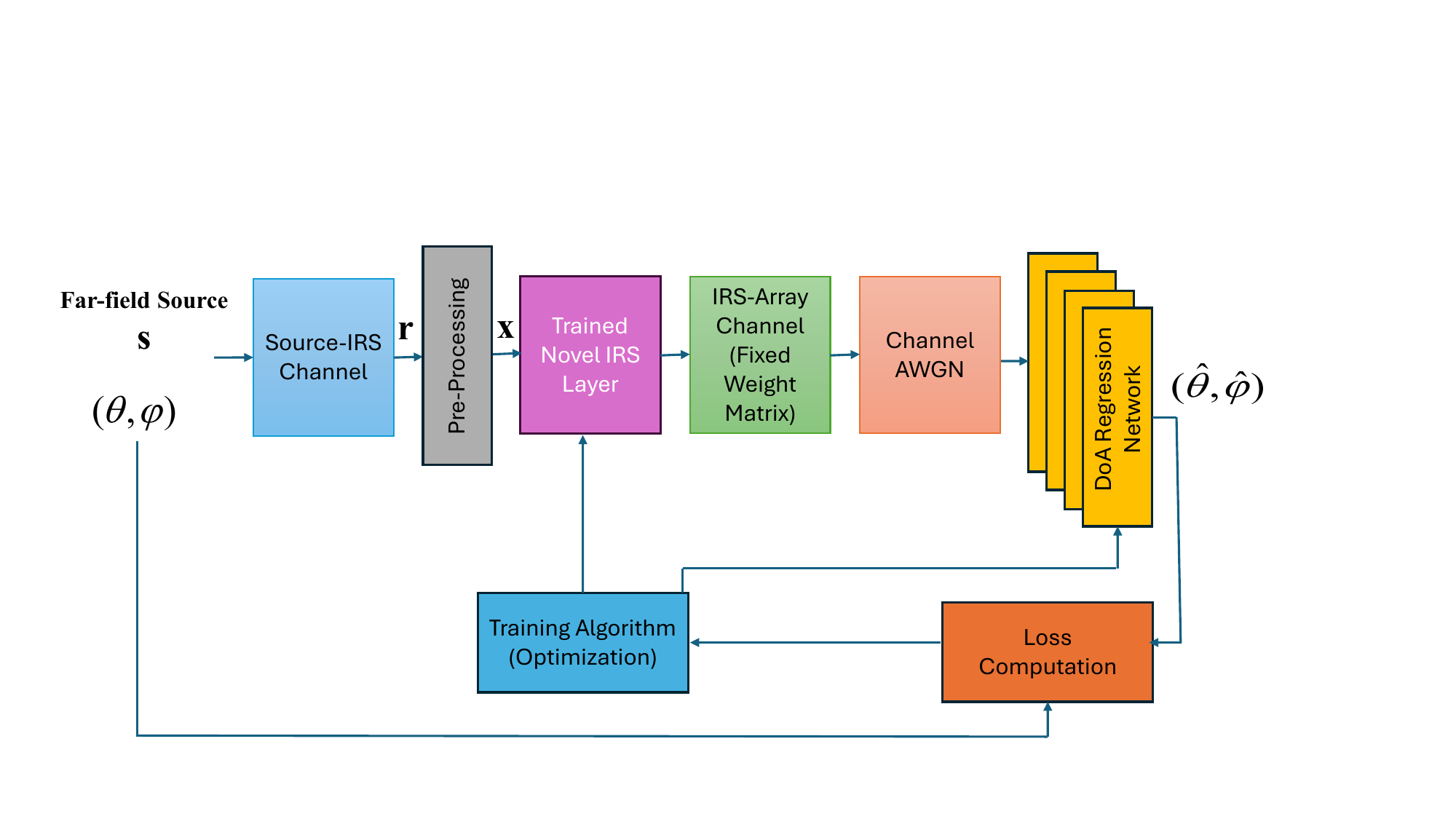}

	\caption{Training}
	\end{subfigure}
	\hfill 
	
	\begin{subfigure}[b]{0.2\textwidth}
	\centering
	
	\hspace*{-25mm}
	\vspace*{0mm}
	\includegraphics[scale=0.3,trim={0cm 4cm 0cm 4cm},clip]{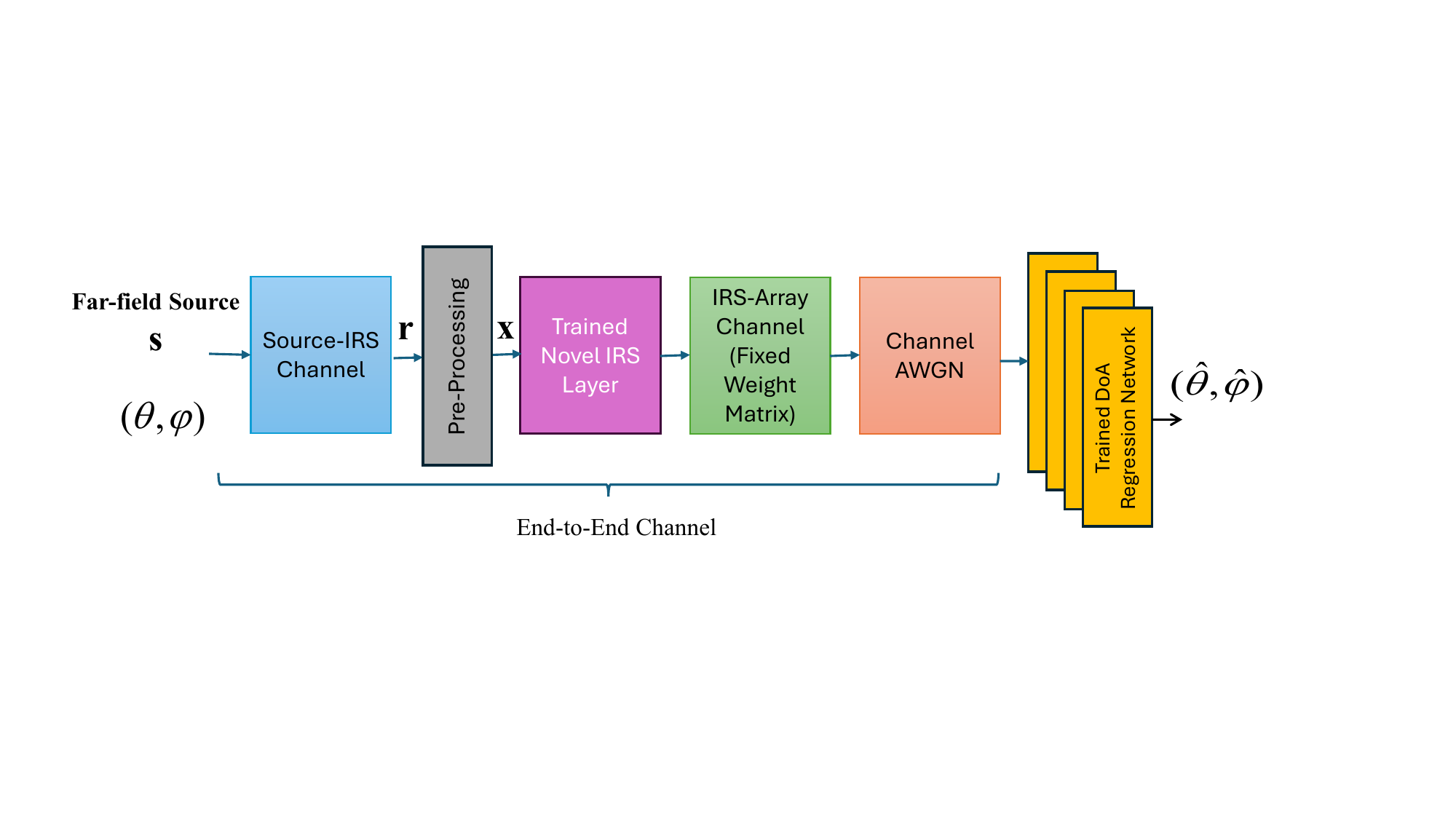}

		\caption{Testing}
	\end{subfigure}
	\caption{The end-to-end IRS-assisted DoA estimation system.}
	\label{fig1005x}
\end{figure}

Subsequently, a DoA regression network is applied to the resulting received signal, i.e., interleaved real and imaginary components of $Y$. The end-to-end system architecture is illustrated in Fig. \ref{fig1005x}.

During the test, Received signal reflected from IRS with trained phases are fed to the DoA regression network, fixed with learned weights from the training phase. 

\subsubsection*{DoA regression networks}

Different architectures, including a proposed structure, are assumed as DoA regressors. Scheme of DoA regressors with corresponding details are illustrated through figures \ref{fig1002a}, \ref{fig1002b}, and \ref{fig1002d}.

In order to have a fair comparison, all structures have approximately the same number of trainable parameters, with 49,253, 49,083, and 48,027 trainable parameters for Fully Connected (FC) network (designed based on \cite{fuchs2019single}), Convolutional Neural Network (CNN) (designed based on \cite{zheng2024deep}), and the proposed structure under 10 snapshots and 50 features per each snapshot (i.e., interleaved real and imaginary parts of $5\times 5$ IRS observations).

The number of nodes corresponding to the output layer is determined based on the number of sources, e.g., equals 2 for the case of one source, representing elevation and azimuth angles.  Due to the fact that input measurements contain both positive and negative values, activation functions are chosen to be $\tanh(\cdot)$ through the network architectures.

\begin{figure}[h]
	\centering
	\resizebox{0.45\textwidth}{!}{
		\begin{tikzpicture}[
			scale=0.45,
			transform shape,
			rotate=90,
			node distance=1.2cm,
			every node/.style={
				draw=black,
				rectangle,
				minimum width=3cm,
				minimum height=0.8cm,
				align=center
			},
			every edge/.style={draw,-{Latex[length=1.5mm]}}
			]
			\definecolor{lightblue}{RGB}{180,200,255}    
			\definecolor{lightgreen}{RGB}{200,255,200}   
			\definecolor{lightyellow}{RGB}{255,255,180}  
			\definecolor{lightcoral}{RGB}{255,182,193}   
			\node[fill=lightblue] (input) {Input \\ (10, 50)};
			\node (flatten) [below of=input, fill=lightgreen] {Flatten};
			\node (dense1) [below of=flatten, fill=lightyellow] {Dense (86, tanh)};
			\node (dropout1) [below of=dense1] {Dropout (0.25)};
			\node (dense2) [below of=dropout1, fill=lightyellow] {Dense (48, tanh)};
			\node (dropout2) [below of=dense2] {Dropout (0.25)};
			\node (dense3) [below of=dropout2, fill=lightyellow] {Dense (32, tanh)};
			\node (dropout3) [below of=dense3] {Dropout (0.5)};
			\node (output) [below of=dropout3, fill=lightyellow] {Dense (2, sigmoid)};
			\path (input) edge (flatten)
			(flatten) edge (dense1)
			(dense1) edge (dropout1)
			(dropout1) edge (dense2)
			(dense2) edge (dropout2)
			(dropout2) edge (dense3)
			(dense3) edge (dropout3)
			(dropout3) edge (output);
		\end{tikzpicture}
	}
	\caption{Architecture of the FC DoA regression network.}
	\label{fig1002a}
\end{figure}

\begin{figure}[h]
	\centering
	\resizebox{0.55\textwidth}{!}{
		\begin{tikzpicture}[
			scale=0.55,
			transform shape,
			rotate=90,
			node distance=1.0cm,
			every node/.style={
				draw=black,
				rectangle,
				minimum width=3cm,
				minimum height=0.8cm,
				align=center
			},
			every edge/.style={draw,-{Latex[length=1.5mm]}}
			]
			\definecolor{lightblue}{RGB}{180,200,255}    
			\definecolor{lightgreen}{RGB}{200,255,200}   
			\definecolor{lightyellow}{RGB}{255,255,180}  
			\definecolor{lightcoral}{RGB}{255,182,193}   
			\node[fill=lightblue] (input) {Input \\ (10, 50)};
			\node (transpose1) [below of=input] {Transpose};
			\node (conv1d1) [below of=transpose1, fill=lightcoral] {Conv1D (24, kernel\_size=3, tanh)};
			\node (dropout1) [below of=conv1d1] {Dropout (0.25)};
			\node (conv1d2) [below of=dropout1, fill=lightcoral] {Conv1D (64, kernel\_size=3, tanh)};
			\node (dropout2) [below of=conv1d2] {Dropout (0.25)};
			\node (conv1d3) [below of=dropout2, fill=lightcoral] {Conv1D (96, kernel\_size=3, tanh)};
			\node (dropout3) [below of=conv1d3] {Dropout (0.25)};
			\node (conv1d4) [below of=dropout3, fill=lightcoral] {Conv1D (64, kernel\_size=3, tanh)};
			\node (dropout4) [below of=conv1d4] {Dropout (0.5)};
			\node (conv1d5) [below of=dropout4, fill=lightcoral] {Conv1D (24, kernel\_size=3, tanh)};
			\node (dropout5) [below of=conv1d5] {Dropout (0.5)};
			\node (transpose2) [below of=dropout5] {Transpose};
			\node (global_avg_pool) [below of=transpose2, fill=lightgreen] {GlobalAveragePooling1D};
			\node (dense1) [below of=global_avg_pool, fill=lightyellow] {Dense (32, tanh)};
			\node (dropout6) [below of=dense1] {Dropout (0.5)};
			\node (output) [below of=dropout6, fill=lightyellow] {Dense (2, sigmoid)};
			\path (input) edge (transpose1)
			(transpose1) edge (conv1d1)
			(conv1d1) edge (dropout1)
			(dropout1) edge (conv1d2)
			(conv1d2) edge (dropout2)
			(dropout2) edge (conv1d3)
			(conv1d3) edge (dropout3)
			(dropout3) edge (conv1d4)
			(conv1d4) edge (dropout4)
			(dropout4) edge (conv1d5)
			(conv1d5) edge (dropout5)
			(dropout5) edge (transpose2)
			(transpose2) edge (global_avg_pool)
			(global_avg_pool) edge (dense1)
			(dense1) edge (dropout6)
			(dropout6) edge (output);
		\end{tikzpicture}
	}
	\caption{Architecture of the Convolutional DoA regression network.}
	\label{fig1002b}
\end{figure}

\begin{figure*}[ht]
	\centering
	\resizebox{0.75\textwidth}{!}{
		\begin{tikzpicture}[
			scale=0.75,
			transform shape,
			rotate=90,
			node distance=0.6cm,
			every node/.style={
				draw=black,
				rectangle,
				minimum width=4.5cm,
				minimum height=1cm,
				align=center
			},
			every edge/.style={draw, -{Latex[length=1.5mm]}, shorten >=1pt}
			]
			\definecolor{lightblue}{RGB}{180,200,255}    
			\definecolor{lightgreen}{RGB}{200,255,200}   
			\definecolor{lightyellow}{RGB}{255,255,180}  
			\definecolor{lightcoral}{RGB}{255,182,193}   
			\definecolor{lightcyan}{RGB}{180,255,255}    
			\node[fill=lightblue] (input) {Input \\ (10, 50)};
			\node[fill=lightcyan] (gru2) [below left=1cm and -1cm of input] {GRU (64)};
			\node[fill=lightcyan] (gru3) [below=of gru2] {GRU (32)};
			\node (dropout1) [below=of gru3] {Dropout (0.5)};
			\node (gru_out) [below=of dropout1] {GRU Output};
			\node (transpose) [below right=1cm and -1cm of input] {Transpose};
			\node[fill=lightcoral] (conv1d2) [below=of transpose] {Conv1D };
			\node[fill=lightcoral] (conv1d3) [below=of conv1d2] {Conv1D };
			\node (dropout2) [below=of conv1d3] {Dropout (0.5)};
			\node (global_avg) [below=of dropout2] {GlobalAveragePooling1D};
			\node (conv_out) [below=of global_avg] {Conv1D Output };
			\node (concatenate) [below=2.75cm of $(gru_out)!0.5!(conv_out)$, fill=lightgreen] {Concatenate };
			\node (dense1) [below=of concatenate, fill=lightyellow] {Dense (64, tanh)};
			\node (dropout3) [below=of dense1] {Dropout (0.5)};
			\node (dense2) [below=of dropout3, fill=lightyellow] {Dense (32, tanh)};
			\node (dropout4) [below=of dense2] {Dropout (0.5)};
			\node (output) [below=of dropout4, fill=lightyellow] {Dense (2, sigmoid)};
			\draw[->] (input.south) -- (gru2.north);
			\draw[->] (gru2.south) -- (gru3.north);
			\draw[->] (gru3.south) -- (dropout1.north);
			\draw[->] (dropout1.south) -- (gru_out.north);
			\draw[->] (input.south) -- (transpose.north);
			\draw[->] (transpose.south) -- (conv1d2.north);
			\draw[->] (conv1d2.south) -- (conv1d3.north);
			\draw[->] (conv1d3.south) -- (dropout2.north);
			\draw[->] (dropout2.south) -- (global_avg.north);
			\draw[->] (global_avg.south) -- (conv_out.north);
			\draw[->] (gru_out.south) -- ++(0,-0.5) -| (concatenate.north);
			\draw[->] (conv_out.south) -- ++(0,-0.25) -| (concatenate.north);
			\draw[->] (concatenate.south) -- (dense1.north);
			\draw[->] (dense1.south) -- (dropout3.north);
			\draw[->] (dropout3.south) -- (dense2.north);
			\draw[->] (dense2.south) -- (dropout4.north);
			\draw[->] (dropout4.south) -- (output.north);
		\end{tikzpicture}
	}
	\caption{Architecture of the Proposed DoA regression Network.}
	\label{fig1002d}
\end{figure*}

	The proposed network architecture (illustrated in Fig. \ref{fig1002d}) effectively captures both temporal and spatial information. Array measurements separate into two parallel branches: a Recurrent Neural Network (RNN) branch using Gated Recurrent Units (GRUs) \cite{cho2014learning} for temporal processing and a Convolutional Neural Network (CNN) branch using 1D convolutional layers for spatial processing:

	\begin{itemize}
		\item \textbf{GRU Branch:} This branch is designed to capture temporal features and dependencies within the data. It comprises two stacked GRU layers. The first {GRU} layer has 64 units and is configured to return the full sequence of outputs. This is followed by a {BatchNormalization} layer. The second {GRU} layer has 32 units and returns only the final output state of the sequence, which is then also followed by a {BatchNormalization} layer. The output of this branch is a feature vector of shape {(batch size, 32)}.
		\item \textbf{Conv1D Branch:} This branch aims to extract local dependencies and features from the data. First, a layer transposes the dimensions of the input, changing its shape from {(batch size, 10, 50)} to {(batch size, 50, 10)}. This transposed data is then processed by two 1D convolutional layers. The first {Conv1D} layer has 64 filters, a kernel size of 5, and uses a hyperbolic tangent activation function, followed by {BatchNormalization}. The second {Conv1D} layer has 32 filters, a kernel size of 3, and also uses {tanh} activation, followed by {BatchNormalization}. Finally, a {GlobalAveragePooling1D} layer reduces the spatial dimension of the output to a single feature vector of shape {(batch size, 32)}, summarizing the learned local patterns.
	\end{itemize}
	
	The outputs from the GRU branch (shape {(batch size, 32)}) and the Conv1D branch (shape {(batch size, 32)}) are then combined using a {Concatenate} layer, resulting in a combined feature vector of shape {(batch size, 64)}.
	
	This combined feature vector is then processed by two dense layers with {tanh} activation, each followed by a {BatchNormalization} layer. {Dropout} layers are applied to prevent overfitting.
	
	Finally, the network concludes with a {Dense} output layer with 2 units and a {sigmoid} activation function (both elevation and azimuth angles are normalized to the range of 0 to 1).
	
	This dual-branch architecture, integrating both recurrent and convolutional processing, allows the model to capture both temporal and spatial features from the input data, potentially leading to more robust and accurate predictions.


Thus the overall IRS-assisted DoA estimation system can be represented with the following transform:

\begin{equation}
	\label{eqn_1019b}
	\hat{\theta} = \text{N}(\mathbf{x}) = \text{DoANET}(\mathbf{W}.\text{R}(\text{P}(\mathbf{x})))+ \mathbf{N'})
\end{equation}
where $\text{P(.)}$ denotes pre-processing, $\text{R(.)}$ represents the IRS layer, and $\mathbf{W}$ represents the fixed weight matrix. $\mathbf{N'}$ denotes the  real noise matrix corresponding the channel. Finally, DoANET(.) represents the network associated with the DoA regression.

The overall system, including both the IRS layer and the DoA estimator network will be tuned and adjusted simultaneously during the training phase concerning the ultimate task of DoA estimation.

Once trained, received signal based on optimized IRS phases are used for DoA estimation. The details and parameters of the network and also the generation procedure of the corresponding datasets are presented in the next section, where the performance of the proposed scheme is compared with other IRS-based DoA estimation methods through extensive simulations.
\section{Numerical Simulations and Results}
\subsection{Setup}
 The number of array elements and the number of IRS unit cells are assumed to be the same and equal to 5 at each dimension (25 in total). The problem model with simulation presets is illustrated in Fig.~\ref{fig1001b}. The source signal is assumed to be a narrow-band deterministic exponential with a frequency of 1 GHz. Since the far-field range includes distances greater than $\frac{2D^2}{\lambda}$ (where $D$ denotes the largest antenna dimension) and respecting that for the IRS and the array, the inter-element spacing is considered to be $\lambda/4$ and $\lambda/2$, respectively, $D = (M^A-1)\lambda/2 = 2.5\lambda = 0.75 \, \text{m}$. Thus, the far-field range in our considered setup will be any distance greater than 3.75 meters. Based on the above far-field range threshold, we have considered the distances mentioned in Fig.~\ref{fig1001b}.
 
 The optimization process for training the neural network employs the Adam optimizer. The initial learning rate is set to 0.015. Training is conducted with a batch size of 64. The model is trained over 50 epochs. To further enhance the training process and prevent overfitting, we use learning rate reduction callback. Moreover, the loss function is assumed to be the MSE.
 
 Initialization for both manifold-based optimization algorithm phases and also the sinusoidal weights phases of the IRS layer is done based on a uniform distribution over the set $[-\pi,\pi]$.
 

\begin{figure}[h]
	\centering
	\resizebox{0.5\textwidth}{!}{%
		\begin{circuitikz}
			\tikzstyle{every node}=[font=\small]
			\draw  (9.5,7.25) -- (8,7.25) -- (8.5,7) -- (10,7) -- cycle;
			\draw [ rotate around={-160:(6,9.75)}] (5.5,9.25) -- (5.75,9.25) -- (6.25,10.5) -- (6,10.5) -- cycle;
			\draw [dashed] (6,7.25) -- (6,7.25);
			\draw [dashed] (8.25,7.25) -- (6,7.25);
			\draw [dashed] (6.25,9) -- (6.25,7);
			\node [font=\small] at (5.5,6.75) {(0,0,0)};
			\node [font=\small] at (7,6.75) {1 m};
			\node [font=\small] at (5.5,8.25) {1 m};
			\draw [ fill={rgb,255:red,5; green,5; blue,5} ] (8,10.75) rectangle (9.25,9.25);
			\draw [->, >=Stealth, dashed] (14.25,10.5) -- (9.25,7);
			\draw [->, >=Stealth, dashed] (9.25,7) -- (6,9.75);
			\node [font=\small] at (12.5,8.5) {10 m};
			\node [font=\small] at (8,7) {IRS};
		\end{circuitikz}
	}%
	\caption{Problem Model with Simulation Presets}
	\label{fig1001b}
\end{figure}

The additive white Gaussian noise matrix \(\mathbf{N'} \in \mathbb{R}^{2M^A \times L}\) is defined as
\begin{equation}
	\mathbf{N'} = \sqrt{\frac{P_r}{2.M^A.L.\text{SNR}}} \mathbf{N}_0,
\end{equation}
where \(P_r\) denotes the power of the received signal at the array, \(\text{SNR}\) is the linear Signal-to-Noise Ratio, and \(\mathbf{N}_0 \in \mathbb{R}^{2M^A \times L}\) is the matrix of independent, zero-mean, unit-variance Gaussian random variables. 

\subsection{DataSets}
For training, we generate a dataset of $N_{train} = 25,000$ examples, where each example is an input-DoA pair $(\mathbf{x}, (\theta, \phi))$. The source direction-of-arrival angles, elevation ($\theta$) and azimuth ($\phi$), are uniformly distributed within the FoV of $0^\circ$ to $90^\circ$ and $0^\circ$ to $180^\circ$, respectively, with a resolution of $0.5^\circ$. The signal-to-noise ratio (SNR) was chosen uniformly from the set \{$-20$, $-10$, 0, 10,20\} dB during training. 20\% of the training set was randomly separated for validation.

The test set consists of $N_{test} = 1000$ examples with random DoAs per each SNR value. The reported performance is the average over these independent test realizations for both classic and learning-based methods.

The root mean squared error (RMSE) defined below, is used as the main performance metric and is plotted against the signal-to-noise ratio (SNR):

\begin{equation}
	\label{eqn_1022}
	\text{RMSE} = \sqrt{\frac{1}{2C} \sum_{c=1}^C ((\hat{\theta}_{c} - \theta)^2 + (\hat{\phi}_{c} - \phi)^2)},
\end{equation}
where $C$ represents the number of Monte Carlo simulations, $\theta, \phi$ denote the actual angular location of the source, and $\hat{\theta}_{c}, \hat{\phi}_{c}$ represent the estimated angular location obtained in the $c$-th simulation run (expressed in degrees). 

\subsection{Results}

\begin{figure*}[h]
	\centering
	
	\begin{subfigure}[b]{0.25\linewidth}
		\centering
		\hspace*{-10mm}
		\vspace*{0mm}
		\includegraphics[scale=0.4,trim={2cm 8cm 3.5cm 8.25cm},clip]{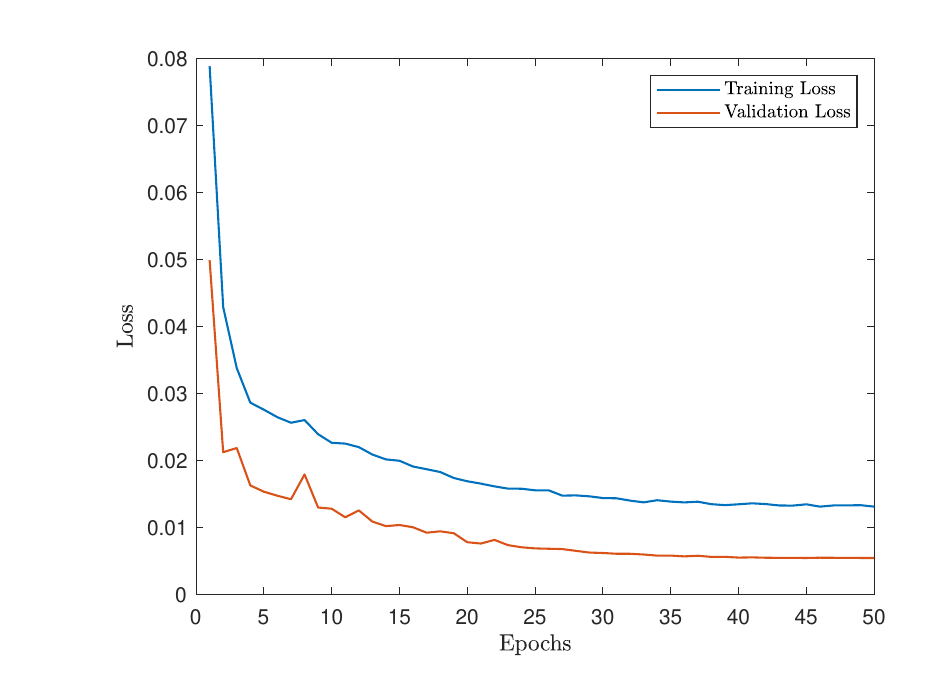}
		\caption{FC Structure}
	\end{subfigure}\hfill
	\begin{subfigure}[b]{0.25\linewidth}
		\centering
		\hspace*{-10mm}
		\vspace*{0mm}
		\includegraphics[scale=0.4,trim={2cm 8cm 3.5cm 8.25cm},clip]{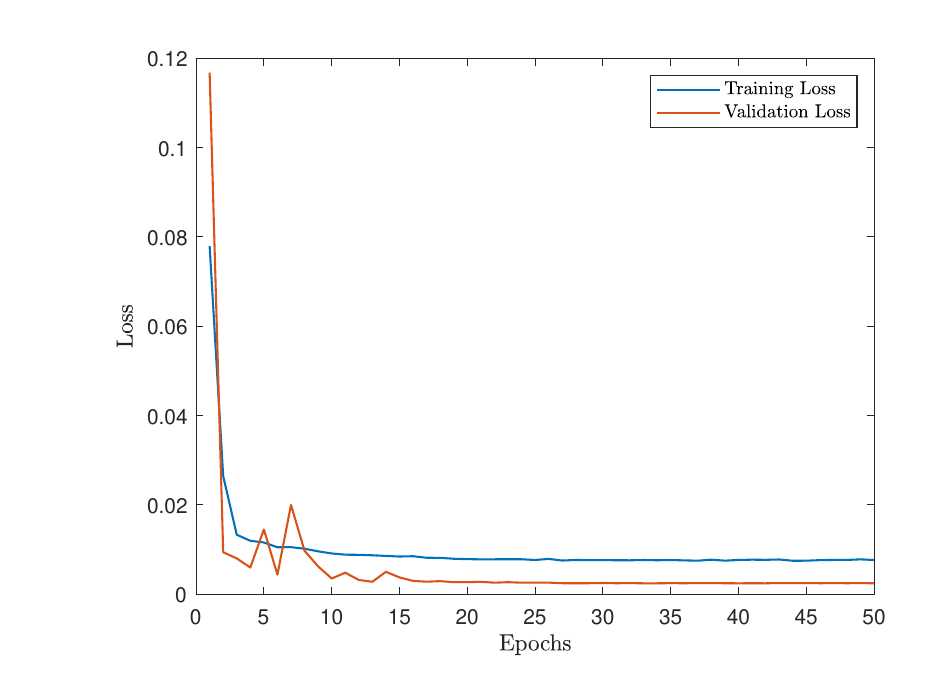}
		\caption{CNN Structure}
	\end{subfigure}\hfill
	\begin{subfigure}[b]{0.25\linewidth}
		\centering
		\hspace*{-10mm}
		\vspace*{0mm}
		\includegraphics[scale=0.4,trim={2cm 8cm 3.5cm 8.25cm},clip]{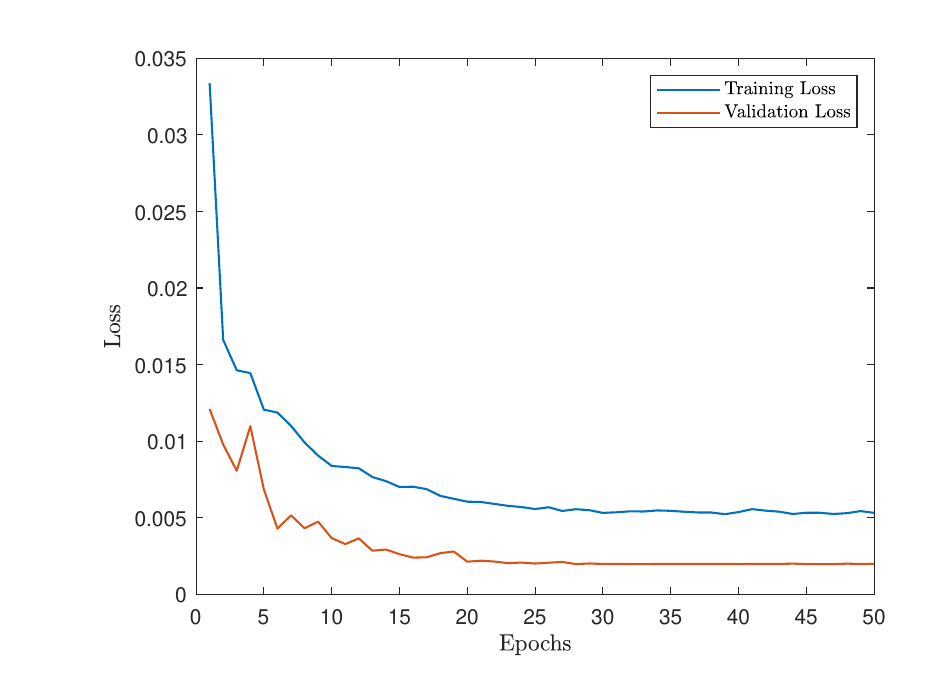}
		\caption{Proposed Structure}
	\end{subfigure}
	\caption{Learning Curves}
	\label{fig:10}
\end{figure*}

We conduct a comparative performance analysis of the proposed learning-based method (with different DoA regression structures) and manifold optimization-based  Maximum Likelihood (ML) direction-of-arrival (DoA) estimation technique. This evaluation is performed assuming 10 snapshots and a randomly located source within the field of view (FoV), over 1000 independent runs per each SNR value. Initially, learning curves corresponding to different DoA regression structures are provided in Fig.~\ref{fig:10}. As observed, since the validation loss is almost always below the training loss, no overfitting occurs with the learning based structures. 

\begin{figure}[H]
	\centering
	
	\hspace*{-10mm}
	\vspace*{0mm}
	\includegraphics[scale=0.6,trim={2cm 8cm 3.5cm 8.25cm},clip]{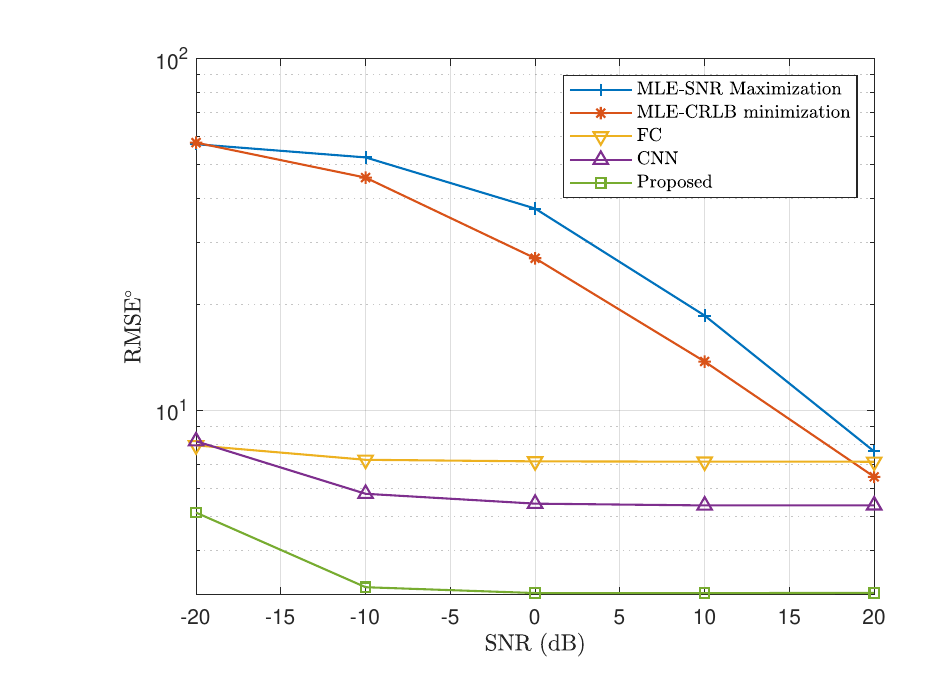}

	\caption{RMSE vs. SNR - 10 Snapshots}
	\label{fig:5}
\end{figure}

The resulting RMSE depicted in Fig.~\ref{fig:5}, is averaged over Monte Carlo runs. It is worth noting that CRLB for the ML approach is derived in prior based on a coarse estimate of the DoA, obtained from ML estimation with randomly initialized IRS phases.
\begin{figure}[H]
	\centering
	
	\hspace*{-10mm}
	\vspace*{0mm}
	\includegraphics[scale=0.6,trim={2cm 8cm 3.5cm 8.25cm},clip]{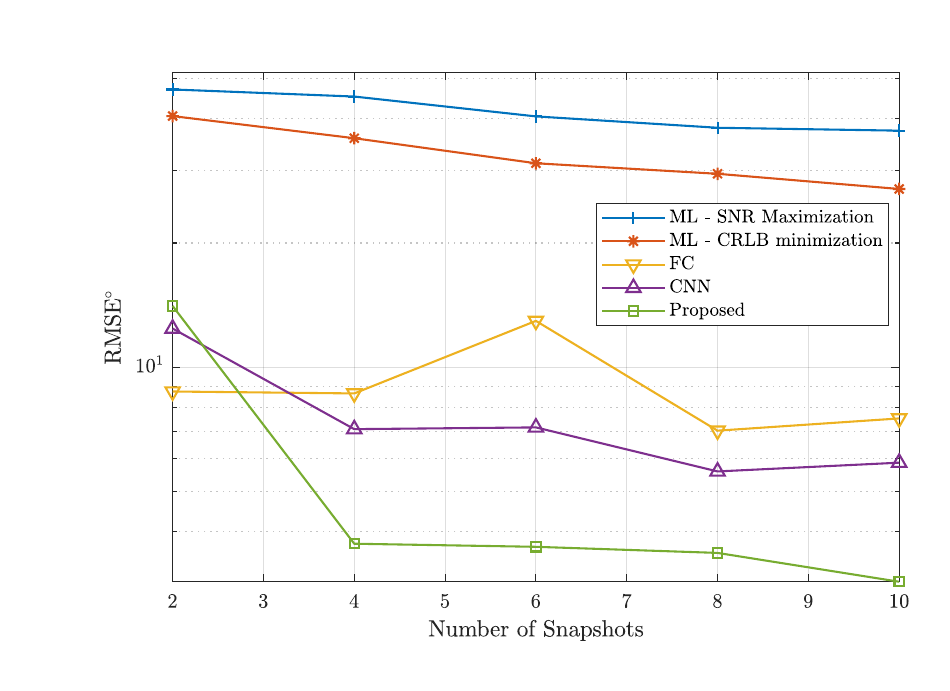}

	\caption{RMSE vs. Number of Snapshots - SNR = $0$ dB}
	\label{fig:201}
\end{figure}
The results demonstrate that proposed learning-based IRS-assisted DoA estimators under different structures achieve superior performance compared to the non-learning methods across the entire range of considered SNR values, which is more obvious under low SNR values. Besides, among learning-based DoA regressor structures, the proposed structure consistently outperforms other approaches. Moreover, under considered conditions, the convolutional structure shows lower RMSE than its fully-connected counterpart.

Due to the fact that IRS-assisted NLoS channels introduce controlled coherent multi-path components, classic approaches, experiences significant performance degradation, which is more obvious for lower SNR values.
 This phenomenon mirrors traditional multi-path coherence challenges but is exacerbated by the deterministic phase-shifting behavior of the IRS, which amplifies signal correlation \cite{du2018direct, zhang2024tdoa}. This indicates another advantage of the proposed learning-based approach compared to the classic methods under such NLoS scenarios. 

Moreover, a key advantage of the learning-based approach is its computational efficiency. Unlike the classical methods, which necessitate performing a corresponding optimization for every randomly positioned test instance, the neural network provides rapid estimations after training.

\begin{figure*}[h]
	\centering
	\begin{subfigure}[b]{0.19\linewidth}
		\centering
		\includegraphics[scale=0.27,trim={2.5cm 7cm 2cm 7cm},clip]{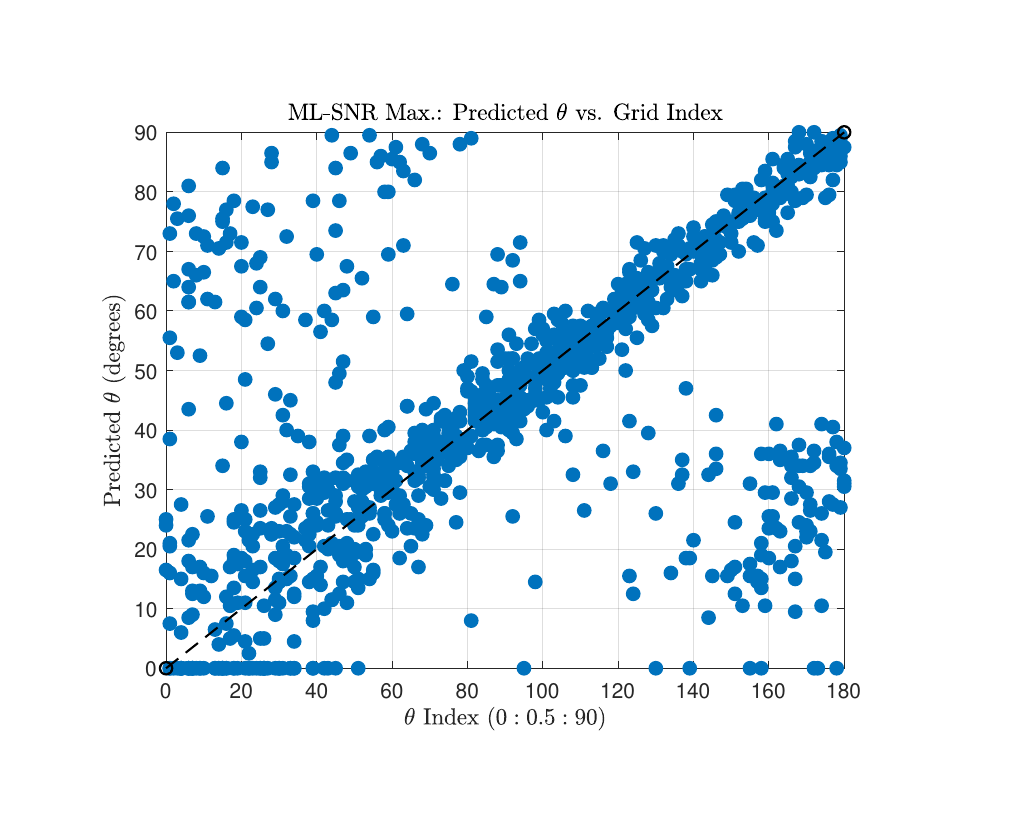}
		\caption{ML-SNR Max.}
	\end{subfigure}\hfill
	\begin{subfigure}[b]{0.19\linewidth}
		\centering
		\includegraphics[scale=0.27,trim={2.5cm 7cm 2cm 7cm},clip]{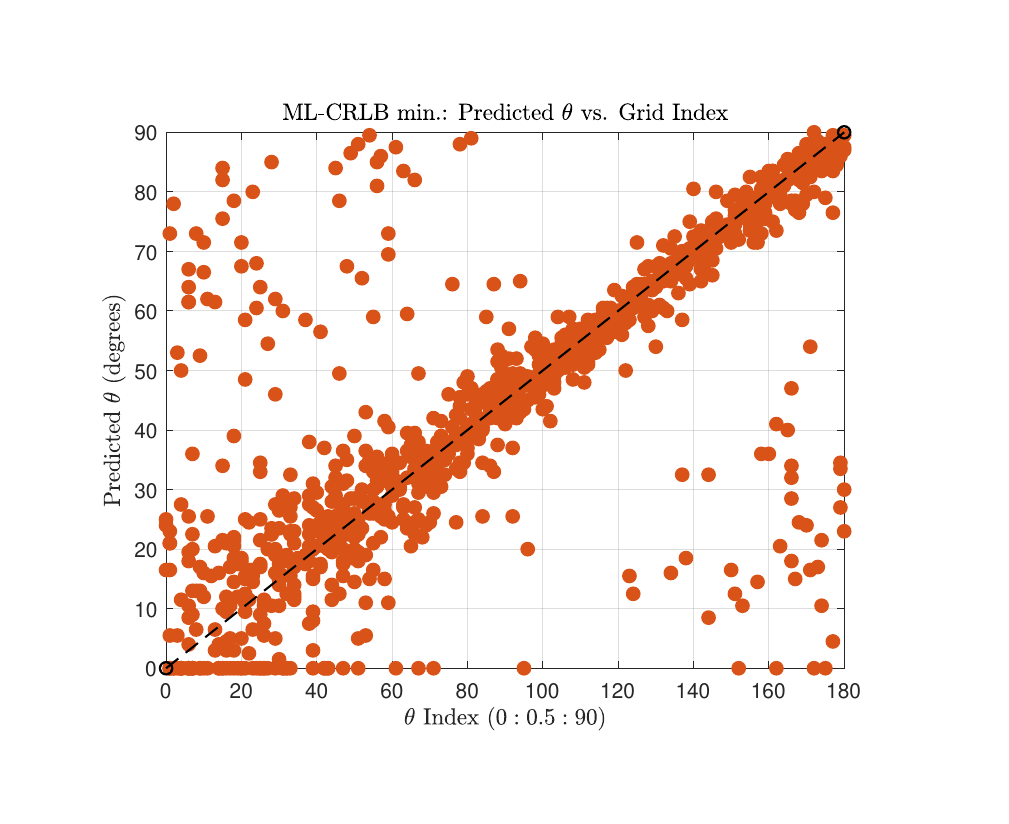}
		\caption{ML-CRLB min.}
	\end{subfigure}\hfill
	\begin{subfigure}[b]{0.19\linewidth}
		\centering
		\includegraphics[scale=0.27,trim={2.5cm 7cm 2cm 7cm},clip]{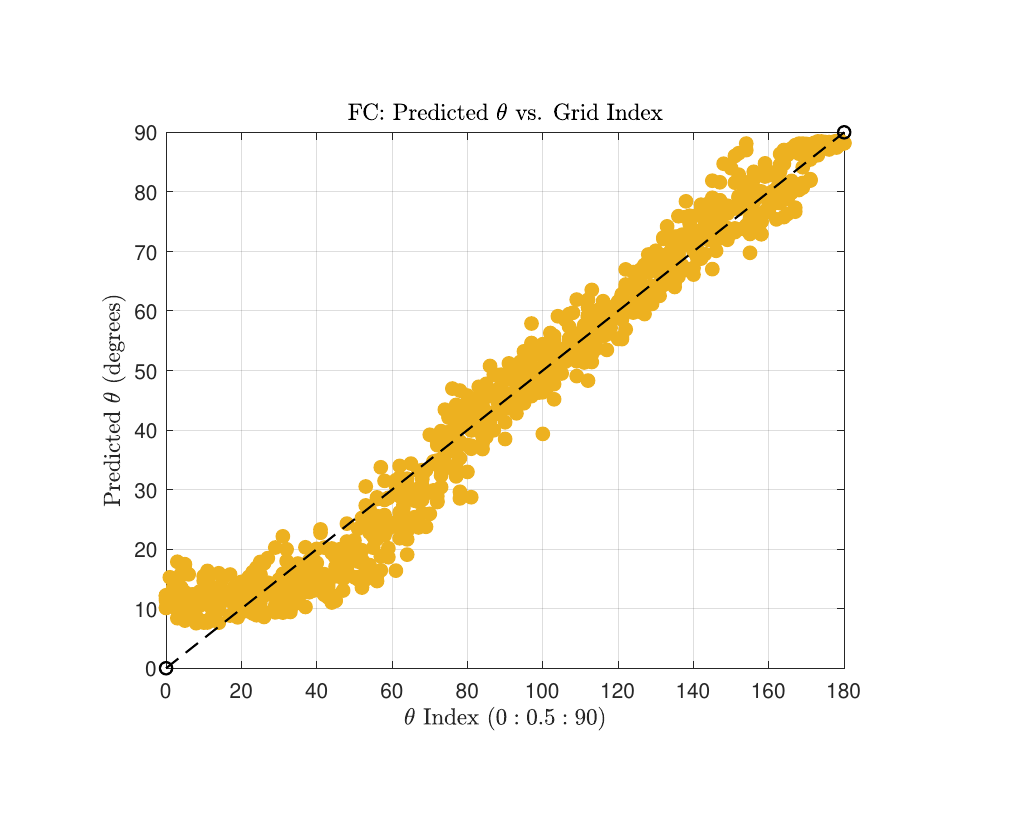}
		\caption{FC}
	\end{subfigure}\hfill
	\begin{subfigure}[b]{0.19\linewidth}
		\centering
		\includegraphics[scale=0.27,trim={2.5cm 7cm 2cm 7cm},clip]{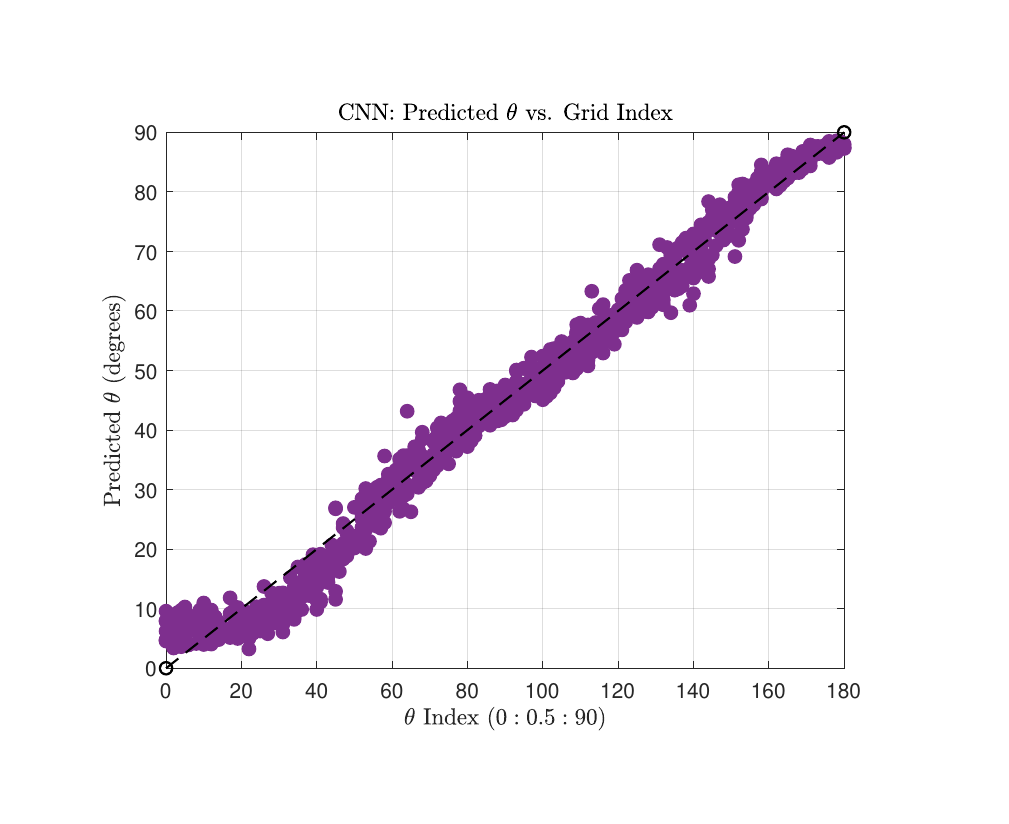}
		\caption{CNN}
	\end{subfigure}\hfill
	\begin{subfigure}[b]{0.19\linewidth}
		\centering
		\includegraphics[scale=0.27,trim={2.5cm 7cm 2cm 7cm},clip]{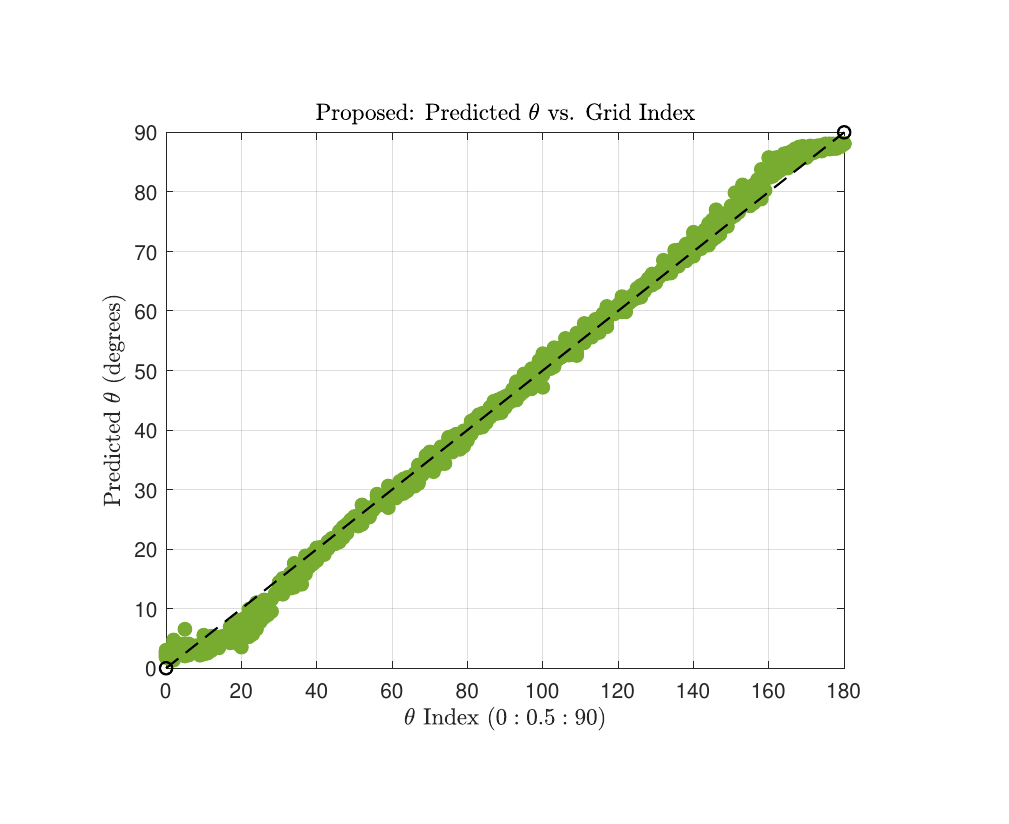}
		\caption{Proposed}
	\end{subfigure}
	
	\vspace{0.1cm}
	
	    \begin{subfigure}[b]{0.19\linewidth}
		\centering
		\includegraphics[scale=0.27,trim={2.5cm 7cm 2cm 7cm},clip]{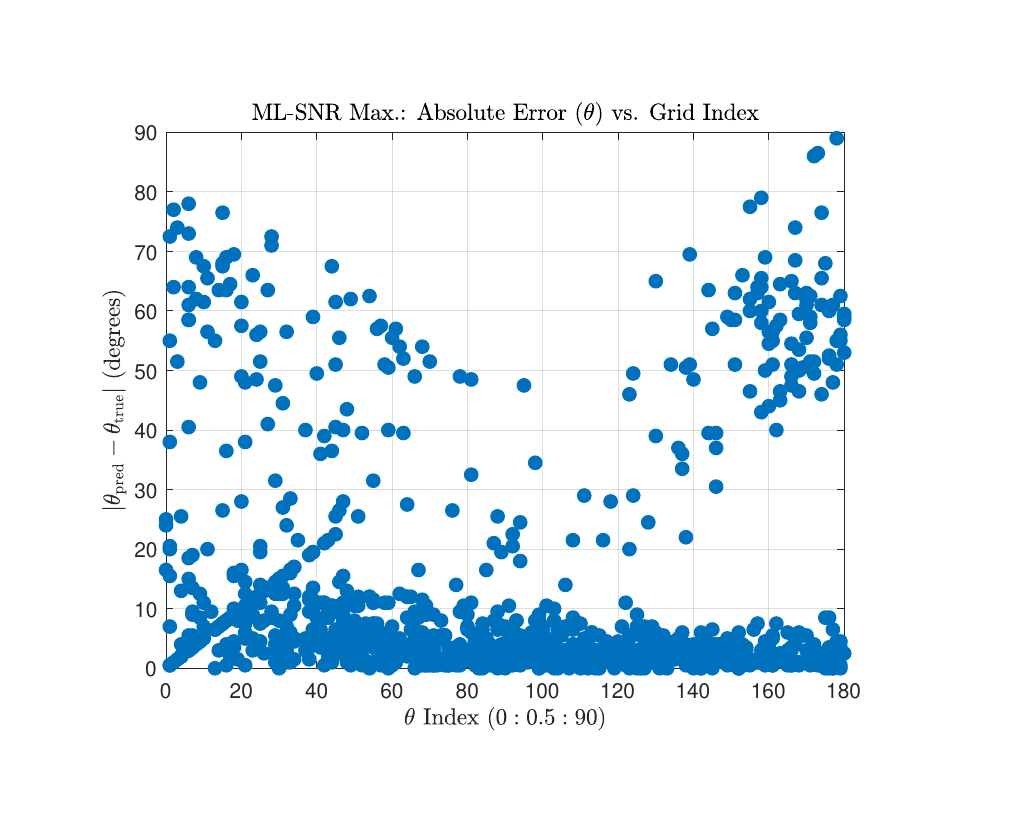}
		\caption{ML-SNR Max.}
		\end{subfigure}\hfill
		\begin{subfigure}[b]{0.19\linewidth}
			\centering
			\includegraphics[scale=0.27,trim={2.5cm 7cm 2cm 7cm},clip]{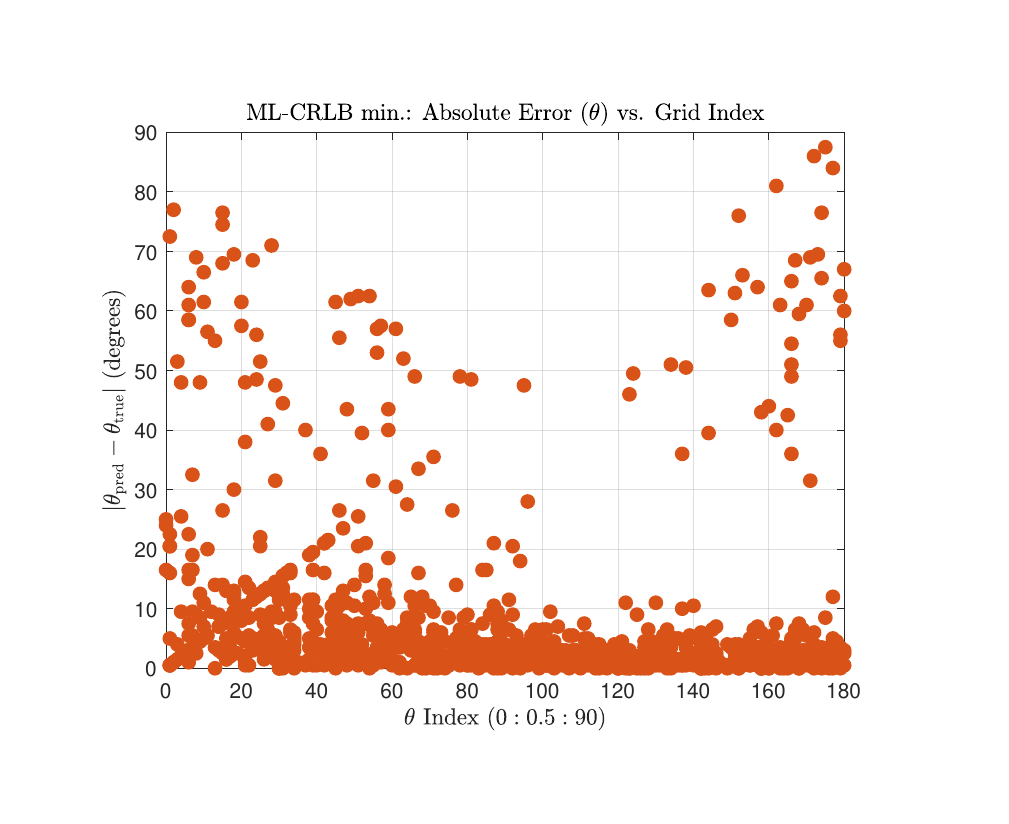}
			\caption{ML-CRLB min.}
		\end{subfigure}\hfill
		\begin{subfigure}[b]{0.19\linewidth}
			\centering
			\includegraphics[scale=0.27,trim={2.5cm 7cm 2cm 7cm},clip]{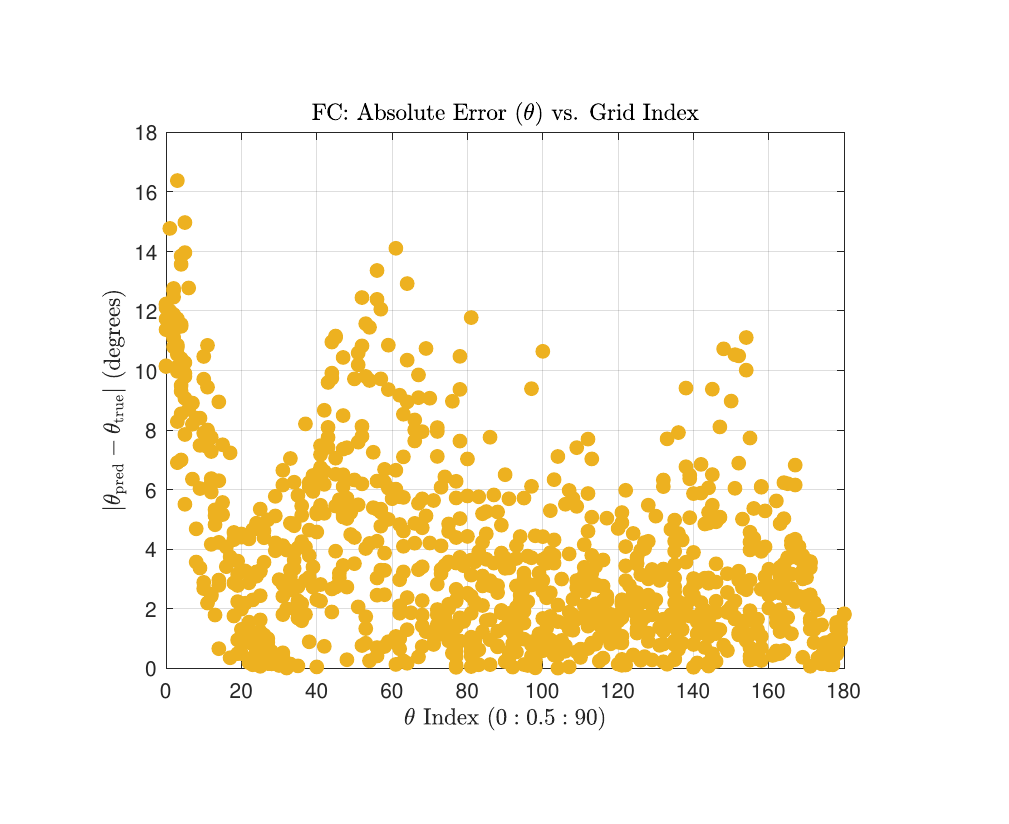}
			\caption{FC}
		\end{subfigure}\hfill
		\begin{subfigure}[b]{0.19\linewidth}
			\centering
			\includegraphics[scale=0.27,trim={2.5cm 7cm 2cm 7cm},clip]{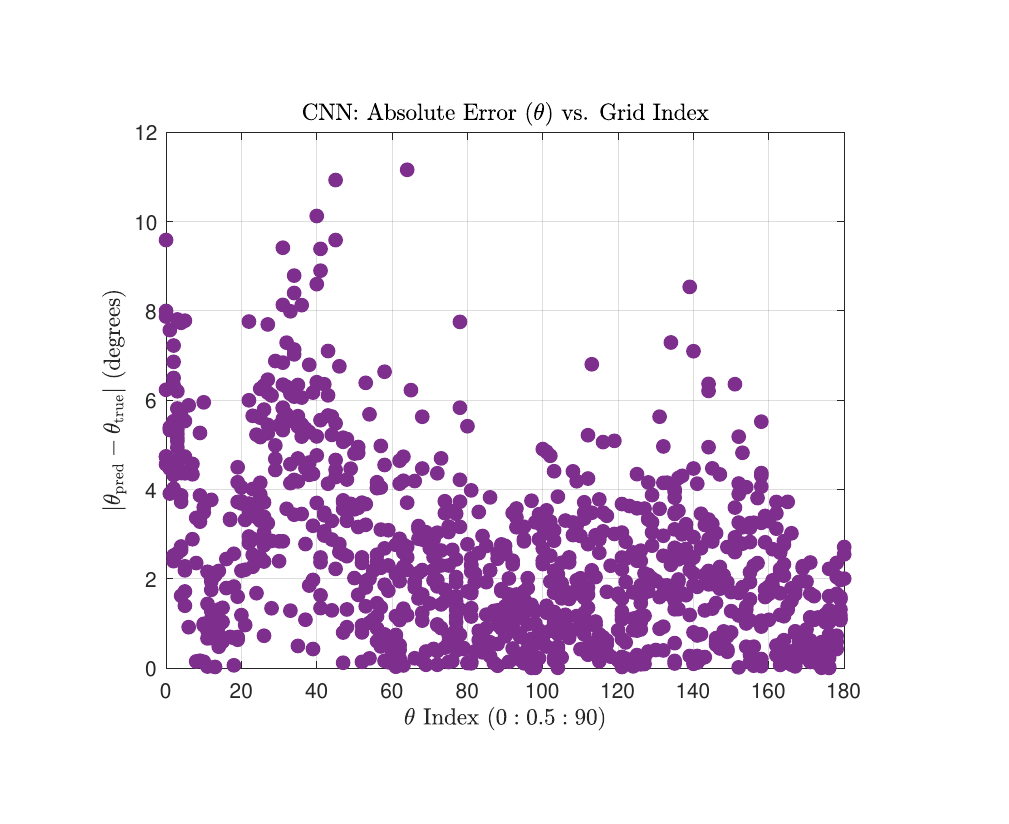}
			\caption{CNN}
		\end{subfigure}\hfill
		\begin{subfigure}[b]{0.19\linewidth}
			\centering
			\includegraphics[scale=0.27,trim={2.5cm 7cm 2cm 7cm},clip]{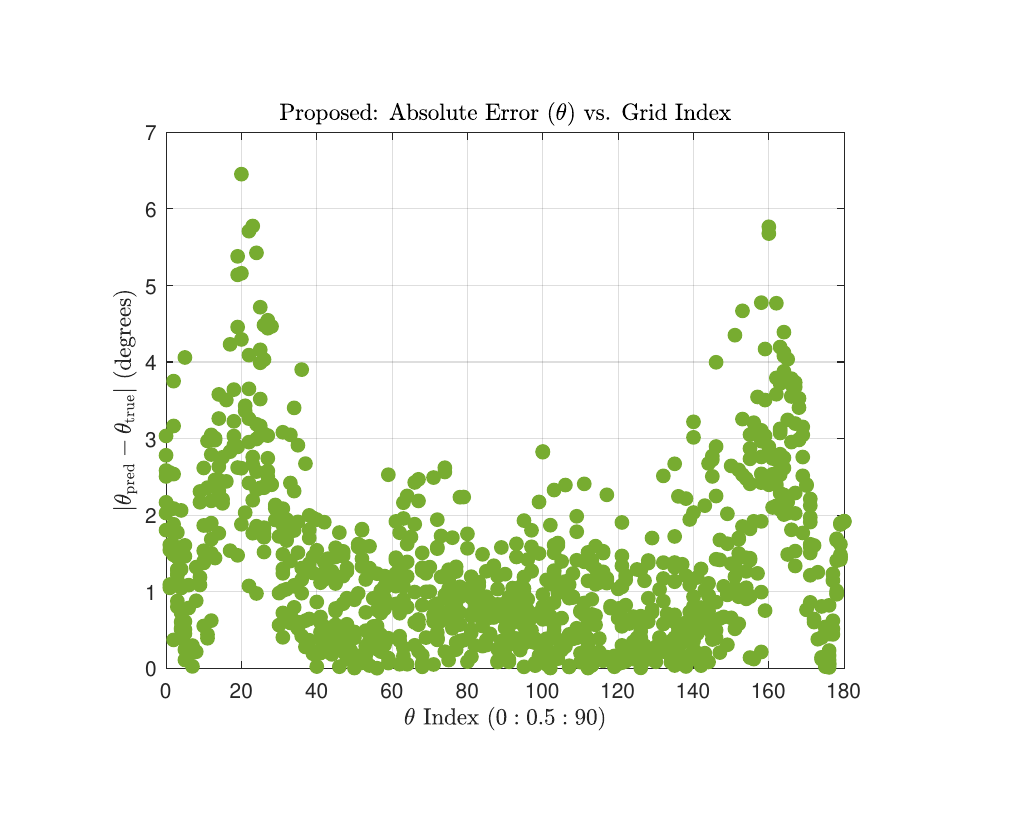}
			\caption{Proposed}
		\end{subfigure}
		
	\caption{Prediction Scatter and Absolute Error Plots for $\theta$ at SNR = 0 dB for Different Methods}
		\label{fig:7_combined}
	\end{figure*}

In the next experiment, we evaluate the performance of the proposed novel learning-based method and other approaches through prediction scatter plot, i.e., predicted angles versus true angle indices, respecting each approach and the ideal condition (represented as a dashed diagonal line). This evaluation is done for $\theta$, assuming 10 snapshots and a randomly located source within the field of view (FoV), under the SNR value of 0 dB and based on 1000 independent runs.
As obseved through figure ~\ref{fig:7_combined}, overall, the learning-based approach based on the proposed DoA regressor outperforms others. Besides, learning-based approaches outperform classic ones under assumed conditions. Additionally, Absolute angular error depicted over the true angle reconfirm these results.
The final experiment assesses the impact of varying the number of snapshots on DOA estimation accuracy under the condition where the SNR value is  $0$ dB. Same as before, a randomly located source with varying number of snapshots is considered. Real-world DoA estimation, especially in dynamic environments often operates with extremely few snapshots due to hardware constraints or fast-moving sources, thus the number of snapshots is assumed to be below 10. For each snapshot count, the RMSE was averaged over 1000 independent Monte Carlo trials. Same neural networks as before were trained using data with corresponding snapshots number. The result, presented in Fig. ~\ref{fig:201}, demonstrate that learning-based approaches maintain superior performance compared to classic ones under these conditions. Besides, the proposed learning-based structure usually outperforms other learning-based methods under different number of snapshots. This highlights the potential of the proposed approach for real-world applications under extreme conditions and with limited data.

\subsection{Computational Complexities}
To evaluate the computational efficiency of the proposed approaches, we analyze the inference complexity of the classic Maximum Likelihood (ML) approaches under Cramér-Rao Lower Bound (CRLB) minimization and Signal-to-Noise Ratio (SNR) maximization criteria, alongside the learning-based approaches (fully connected (FC), convolutional neural network (CNN), and proposed) for DoA estimation. The complexity is measured in floating-point operations (FLOPs) per sample, assuming an array with \( M_A = 25 \) antennas, \( M_R = 25 \) IRS elements, and \( L = 10 \) snapshots. The system operates with a field of view of \( \theta \in [0^\circ, 90^\circ] \) (grid size \( G_\theta = 181 \), 0.5-degree resolution) and \( \phi \in [0^\circ, 180^\circ] \) (grid size \( G_\phi = 361 \), 0.5-degree resolution).

For the ML approach under CRLB minimization, the complexity comprises two stages: (1) Riemannian optimization of IRS phase shifts using the Manopt \cite{manopt} trust-regions solver on a complex circle manifold, and (2) DoA estimation via grid search. The Riemannian optimization involves computing the cost function, gradient, and Hessian approximation, dominated by matrix operations of \( O(M_A \cdot M_R) \), involving the channel matrix (\( M_A \times M_R \)) and multiple evaluations per iteration. The DoA estimation dominates the overall complexity due to the exhaustive grid search over \( G_\theta \times G_\phi = 65,341 \) points.

For the ML approach under SNR maximization, the optimal phase for the \( n \)-th IRS unit cell is designed according to (\ref{eqn_1010}).
This phase design simplifies the optimization by directly computing phase shifts based on geometric distances, reducing the need for iterative optimization. However, the DoA estimation still requires a grid search over \( G_\theta \times G_\phi \), maintaining high computational complexity.

The learning-based approaches perform a single forward pass through neural networks, processing complex-valued inputs (\( M_A \times L = 25 \times 10 \)) split into 25 real and 25 imaginary parts, forming a \( (10, 50) \) input tensor. Neural network inference was conducted with a batch size of 64, consistent with training, to leverage computational efficiency. To ensure a fair comparison with ML approaches, which process single samples, we report per-sample FLOPs by dividing batch FLOPs by 64. The FC model uses Dense layers with 86, 48, 32, and 2 neurons. The CNN model employs five Conv1D layers with 24, 64, 96, 64, and 24 filters (kernel size 3) followed by a GlobalAveragePooling1D layer and Dense layers with 32 and 2 neurons. The proposed model combines GRU layers (64 and 32 units) with Conv1D layers (64 and 32 filters, kernel sizes 5 and 3) and Dense layers (64, 32, and 2 neurons). FLOPs are calculated by summing the contributions of all relevant layers, using exact layer-specific input sizes (\( D_i \), \( D_j \), \( D_k \), \( C_{j,in} \), \( I_k \), \( T_k \)).

For neural methods, complexity formulas use \( N_i \), the number of neurons in Dense layers; \( F_j \), the number of filters in Conv1D layers; \( K_j \), the kernel size in Conv1D layers; \( C_{j,in} \), the number of input channels to Conv1D layers; \( U_k \), the number of units in GRU layers; \( I_k \), the input feature size to GRU layers; \( T_k \), the number of timesteps in GRU layers; and \( D_i \), \( D_j \), \( D_k \), the input sizes to Dense, Conv1D, and GRU layers, respectively. For ML approaches, the complexity is dominated by the grid search and, for CRLB minimization, the Riemannian optimization iterations (\( I = 100 \)). Table~\ref{tab:complexity} summarizes the per-sample FLOPs and complexity formulas. The ML approaches require approximately \( 10^9 \) FLOPs, driven by the grid search, while neural networks require \( 10^3 \) to \( 10^4 \) FLOPs, making them 4–5 orders of magnitude more efficient. The higher complexity of the proposed approach compared to its FC counterpart is justified by its superior RMSE performance (see Fig.~\ref{fig:5}), reflecting a trade-off for enhanced accuracy.

\begin{table}[!ht]
	\centering
	\caption{Complexity of DoA Estimation (Per Sample). Grid: $G_\theta=181$, $G_\phi=361$. Parameters: $N_i$ (86, 48, 32, 2 for FC; 32, 2 for CNN; 64, 32, 2 for Proposed); $F_j$ (24, 64, 96, 64, 24 for CNN; 64, 32 for Proposed); $K_j$ (3 for CNN; 5, 3 for Proposed); $C_{j,in}$ (10, 24, 64, 96, 64 for CNN; 10, 64 for Proposed); $U_k$ (64, 32 for Proposed); $I_k$ (50, 64); $T_k$ (10, 10); $D_i$, $D_j$, $D_k$ as inputs.}
	\label{tab:complexity}
	\footnotesize
	\setlength{\tabcolsep}{0.5pt}
	\begin{tabular}{l c c}
		\toprule
		Method & FLOPs Per Sample & Complexity Formula \\
		\midrule
		ML (CRLB min.) & $9 \times 10^8$ & $O(M_A^2 L + G_\theta G_\phi M_A^2 + I M_A M_R)$ \\
		ML (SNR Max.) & $9 \times 10^8$ & $O(M_A^2 L + G_\theta G_\phi M_A^2 + M_A M_R)$ \\
		FC & $1.5 \times 10^3$ & $O\left( \sum_{i=1}^4 D_i N_i \cdot 2 \right)$ \\
		\multirow{2}{*}{CNN} & \multirow{2}{*}{$6.5 \times 10^4$} & $O\left( \sum_{j=1}^5 D_j F_j K_j C_{j,in} \cdot 2 \right.$ \\
		& & $\left. + \sum_{i=1}^2 D_i N_i \cdot 2 \right)$ \\
		\multirow{2}{*}{Proposed} & \multirow{2}{*}{$1.95 \times 10^4$} & $O\left( \sum_{k=1}^2 T_k U_k (3 U_k + I_k) \cdot 2 \right.$ \\
		& & $\left. + \sum_{j=1}^2 D_j F_j K_j C_{j,in} \cdot 2 + \sum_{i=1}^3 D_i N_i \cdot 2 \right)$ \\
		\bottomrule
	\end{tabular}
\end{table}

\section{Conclusion}
This paper proposes a novel neural network-based solution for IRS-assisted DoA estimation under NLoS scenarios, centered around a novel trainable IRS layer. The end-to-end system architecture is carefully designed to incorporate the necessary considerations in modeling the problem, enabling the network to effectively learn and adjust the IRS phase configurations. This integrated approach unifies the processes of IRS phase adjustment and DoA estimation, eliminating the need for separate, often complex, optimization algorithms. Numerical simulations across various scenarios confirm the effectiveness and robustness of the proposed method, demonstrating superior performance compared to existing non-learning DoA estimation techniques in complex IRS-assisted NLoS environments under controlled coherent multipath components. Moreover, the proposed learning-based approach (under different implementations) also exhibits lower computational complexity than non-learning ones.
\FloatBarrier

\bibliographystyle{IEEEtran}
\bibliography{MSc_based_refrences_1}


%








\end{document}